\documentclass[12pt,english,floatfix,superscriptaddress,aps,prd,preprint,nofootinbib]{revtex4}
\usepackage{amsmath}
\usepackage{amssymb}
\usepackage{amsbsy}
\usepackage[a4paper, margin=1.8cm]{geometry}
\usepackage{amsfonts}
\usepackage{amsopn}
\usepackage{amstext}
\usepackage{graphicx}
\usepackage{amssymb}
\usepackage{amsfonts}
\usepackage{amsmath}
\usepackage{graphicx}
\usepackage[english]{babel}
\usepackage{color}
\usepackage{xcolor}
\usepackage{slashed}
\usepackage{esint}
\usepackage[dvips]{epsfig}
\usepackage[dvips]{graphicx}
\usepackage{float}
\usepackage{units}
\usepackage{textcomp}
\usepackage{placeins}
\usepackage{hyperref}             % Enable clickable links
\hypersetup{
    colorlinks=true,              % Colored links instead of boxed
    breaklinks=true,              % Allow links to break across lines
    citecolor=blue,               % Color for citation links
    linkcolor=[rgb]{0,0.5,0.9},   % Color for internal links
    urlcolor=red,                 % Color for URLs
    filecolor=green               % Color for file links
}

\usepackage{hyperref}
\usepackage{slashed}

\newcommand{\ie}{\begin{equation}}
\newcommand{\fe}{\end{equation}}
 \newcommand{\bq}{\begin{equation}}
 \newcommand{\eq}{\end{equation}}
 \newcommand{\bqn}{\begin{eqnarray}}
 \newcommand{\eqn}{\end{eqnarray}}
 \newcommand{\nb}{\nonumber}
 \newcommand{\lb}{\label}

\begin{document}
%%%%%%%%%%%%%%%%%%%%%%%%%%%%%%%%%%%%%%%%%%%%%%%%%%%%%%%%%%%%%%%%%%%%%%%%

\title{Propagation effects of Lorentz violation in gravitational waves}

%%%%%%%%%%%%%%%%%%%%%%%%%%%%%%%%%%%%%%%%%%%%%%%%%%%}

%%%%%%%%%%%%%%%%%%%%%%%%%%%%%%%%%%%%%%%%%%%%%%%%%%%%%%%%%%%%%%%%%%%%%%

\author{A. A. Ara\'{u}jo Filho}
\email{dilto@fisica.ufc.br}
\affiliation{Departamento de Física, Universidade Federal da Paraíba, Caixa Postal 5008, 58051--970, João Pessoa, Paraíba,  Brazil.}
\affiliation{Departamento de Física, Universidade Federal de Campina Grande Caixa Postal 10071, 58429-900 Campina Grande, Paraíba, Brazil.}
\affiliation{Center for Theoretical Physics, Khazar University, 41 Mehseti Street, Baku, AZ-1096, Azerbaijan.}
%%%%%%%%%%%%%%%%%%%%%%%%%%%%%%%%%%%%%%%%%%%%%%%%%%%%%%%%%%%%%%%%%%%%%%%%%%%%%%%%%%%%%%%%%%%%%%%%%%%%%%%%%%%%%%%%%%%%%%%%%%%%%%%%%%%%%%%%%%%%%%%%%%%%%%%%%%%%%%%%%%%%%%%%%%%%%%%%%%%%%%%%%%%%%%%%%%%%%%%%%%%%%%%%%%%%%%%%%%%%%%%%%%%%%%%%%%%%%%%%%%%%%%%%%%%%%%%%%%%%%%%%%%%%%%%%%%%%%%%%%%%%
\author{N. Heidari}
\email{heidari.n@gmail.com}

\affiliation{Departamento de Física, Universidade Federal de Campina Grande Caixa Postal 10071, 58429-900 Campina Grande, Paraíba, Brazil.}
\affiliation{Center for Theoretical Physics, Khazar University, 41 Mehseti Street, Baku, AZ-1096, Azerbaijan.}
\affiliation{School of Physics, Damghan University, Damghan, 3671641167, Iran.}

%%%%%%%%%%%%%%%%%%%%%%%%%%%%%%%%%%%%%%%%%%%%%%%%%%%%%%%%%%%%%%%%%%%%%%%%%%%%%%%%%%%%%%%%%%%%%%%%%%%%%%%%%%%%%%%%%%%%%%%%%%%%%%%%%%%%%%%%%%%%%%%%%%%%%%%%%%%%%%%%%%%%%%%%%%%%%%%%%%%%%%%%%%%%%%%%%%%%%%%%%%%%%%%%%%%%%

\author{Iarley P. Lobo}
\email{lobofisica@gmail.com}

\affiliation{Departamento de Física, Universidade Federal de Campina Grande Caixa Postal 10071, 58429-900 Campina Grande, Paraíba, Brazil.}
\affiliation{Department of Chemistry and Physics, Federal University of Para\'iba, Rodovia BR 079 - km 12, 58397-000 Areia-PB,  Brazil.}

%%%%%%%%%%%%%%%%%%%%%%%%%%%%%%%%%%%%%%%%%%%%%%%%%%%%%%%%%%%%%%%%%%%%%%%
\date{\today}

\begin{abstract}

We investigate the propagation of gravitational waves in the presence of Lorentz-- and diffeomorphism--violating operators within the linearized gravitational sector of the Standard Model Extension. Focusing on isotropic contributions, we analyze the combined effects of the CPT--even dimension--four coefficient $\mathring{k}^{(4)}_{(I)}$ and the CPT--odd dimension--five coefficient $\mathring{k}^{(5)}_{(V)}$ on tensorial gravitational radiation. The modified dispersion relation induces both a rescaling of the propagation speed and helicity--dependent corrections, leading to birefringence and polarization mixing without introducing additional propagating degrees of freedom. We derive the retarded Green function associated with the modified wave operator and obtain explicit expressions for the gravitational waveform generated by matter sources. As an application, we examine a binary black hole system and show how Lorentz violation alters the observed strain through shifted retarded times, amplitude rescaling, and higher derivative corrections to the quadrupole formula. 
{Using GW170817/GRB 170817A, published GWTC--3 propagation tests, and conservative polarization consistency arguments, we translate existing observational constraints into bounds on $\mathring{k}^{(4)}_{(I)}$ and $\mathring{k}^{(5)}_{(V)}$.}
\end{abstract}


\maketitle
\tableofcontents

\section{Introduction}

General Relativity has established itself as the prevailing geometric framework for gravitational interactions, having undergone extensive scrutiny across a wide range of physical regimes. Its predictions have been confronted with high–precision measurements in laboratory experiments, solar–system tests, binary pulsars, gravitational–wave observations, and cosmological probes, consistently showing remarkable agreement with observational data \cite{Manchester:2015mda,Will:1997bb,Hoyle:2000cv,Jain:2010ka,Clifton:2011jh,Koyama:2015vza,Stairs:2003eg,Wex:2014nva,Kramer:2016kwa,Berti:2015itd}. Despite this empirical success, the theory does not constitute a complete description of gravitation.
Several conceptual and theoretical obstacles remain unresolved within the classical formulation. At short distances, the occurrence of spacetime singularities signals a breakdown of the theory, while attempts to embed gravity into a consistent quantum framework encounter persistent difficulties. 

On the other hand, on cosmological and galactic scales, the dynamics inferred from observations require additional ingredients—typically modeled as dark matter and dark energy—that do not arise naturally from the Einstein equations themselves. These shortcomings suggest that the classical picture may represent only an effective description of a more fundamental gravitational theory.
Such considerations have stimulated the development of alternative frameworks that extend or modify General Relativity. A number of these proposals are rooted in ideas from quantum gravity, including string–inspired constructions \cite{Kostelecky:1988zi,Kostelecky:1991ak}, loop–based quantization schemes \cite{Gambini:1998it}, and scenarios involving extra dimensions or brane embeddings \cite{Burgess:2002tb}. A common feature shared by many of these approaches is a departure from one of the foundational assumptions of General Relativity: the exact and universal validity of Lorentz invariance and diffeomorphism symmetry.

An effective field theory perspective offers a natural arena for investigating potential violations of Lorentz invariance and diffeomorphism symmetry in gravitational physics. Within this approach, the Standard Model Extension (SME) was introduced as a comprehensive parametrization scheme designed to capture symmetry--breaking effects without committing to a specific microscopic theory \cite{Colladay:1996iz,Kostelecky:2003fs}. The construction of the SME does not rely on altering the geometric foundations of spacetime. Instead, it augments the action by systematically including all operators compatible with observer covariance that violate local Lorentz and diffeomorphism symmetries, classified according to their mass dimension and tensorial structure \cite{Liu:2025bpp}.

Over the past decades, this framework has played a central role in empirical searches for Lorentz violation, particularly in the nongravitational sector, where it has supported an extensive program of experimental and observational tests. The resulting bounds span a wide range of physical systems and energy scales. Once gravitational interactions are incorporated, the SME provides a unified language for analyzing precision measurements drawn from diverse observational settings. These include lunar laser ranging experiments \cite{Bourgoin:2017fpo,Battat:2007uh}, atom-interferometry tests \cite{Muller:2007es}, studies of ultra--high--energy cosmic rays \cite{Kostelecky:2015dpa}, pulsar--timing data \cite{Shao:2014oha,Shao:2019tle,Shao:2014bfa,Shao:2019cyt,Jennings:2015vma,Shao:2018vul}, analyses of planetary ephemerides \cite{Hees:2015mga}, and measurements performed with superconducting gravimeters \cite{Flowers:2016ctv}. In the majority of these investigations, observable effects arise predominantly through modified couplings between gravitational fields and matter degrees of freedom \cite{Kostelecky:2010ze}.

An alternative viewpoint is obtained by isolating the purely gravitational sector and examining its intrinsic dynamics. In this setting, Lorentz-- and diffeomorphism--violating contributions are probed directly through their impact on the propagation of gravitational disturbances. Adopting a linearized description of the metric, gravitational waves provide a clean and sensitive channel for testing departures from standard relativistic behavior, independent of matter couplings \cite{Bailey:2006fd,Kostelecky:2017zob,Ferrari:2006gs}.

In a weak--field description where gravity is expanded about a Minkowski background, possible violations of Lorentz symmetry enter through additional terms in the quadratic action for metric perturbations. These corrections are naturally classified within an effective field–theory expansion by their mass dimension $d$. A further distinction emerges once their behavior under linearized coordinate transformations is taken into account. Specifically, the transformation $h_{\mu\nu} \rightarrow h_{\mu\nu} + \partial_\mu \xi_\nu + \partial_\nu \xi_\mu$, with $h_{\mu\nu}$ representing small deviations from flat spacetime and $\xi_\mu$ an arbitrary infinitesimal vector field, provides the criterion for identifying whether a given operator preserves or violates diffeomorphism symmetry at the linearized level.

Imposing invariance under this gauge transformation severely restricts the allowed operator content. In particular, Lorentz--violating terms that remain compatible with linearized diffeomorphism symmetry first appear at mass dimension $d \ge 4$ \cite{Kostelecky:2017zob,Kostelecky:2016kfm}. The consequences of such higher--dimensional operators for gravitational wave propagation, together with the corresponding phenomenological constraints, have been analyzed extensively in the literature \cite{Kostelecky:2016kfm,Hou:2024xbv,Dong:2023nau,Zhu:2023rrx,Mewes:2019dhj,Kostelecky:2017zob,Haegel:2022ymk,ONeal-Ault:2021uwu,Bailey:2023lzy,Nilsson:2023szw,AouladLafkih:2025stw,Arzano:2016twc}.

If the requirement of diffeomorphism invariance is instead relaxed in the linearized theory, the operator spectrum broadens substantially. In this case, Lorentz--violating contributions with lower mass dimension become admissible, including terms with $d=2$ and $d=3$ \cite{Kostelecky:2017zob}. Such operators introduce qualitatively different modifications to the dynamics of metric perturbations and open additional possibilities for departures from standard relativistic propagation.

The systematic construction of Lorentz--violating contributions to the quadratic action for metric perturbations has been carried out in a comprehensive manner in Ref.~\cite{Kostelecky:2017zob}. That analysis identifies all admissible operators within the linearized gravitational sector, encompassing both terms that remain compatible with infinitesimal coordinate invariance and those that explicitly violate it. Once such operators are introduced, the propagation of gravitational waves departs from the standard relativistic picture in a generic way. Modifications typically appear in the form of anisotropic propagation, helicity--dependent phase velocities, and dispersive time delays, each of which can be encoded directly in the waveform morphology.

When linearized diffeomorphism symmetry is retained, the resulting corrections to gravitational wave dynamics admit a particularly transparent formulation. In this case, the altered propagation laws and their observational consequences have been derived explicitly in Ref.~\cite{Mewes:2019dhj}. These results provide a direct bridge to data analysis, allowing the Lorentz--violating coefficients of the gravitational SME to be constrained through Bayesian inference applied to gravitational wave observations. A related but conceptually distinct approach has been developed using a parametrized framework tailored to symmetry--breaking effects in gravitational wave propagation \cite{Zhu:2023rrx}. Rather than starting from a specific operator basis, this method characterizes deviations at the level of effective propagation parameters. This formalism has been employed to investigate possible Lorentz--violating features in the primordial gravitational wave background, as explored in Ref.~\cite{Li:2024fxy}.

In this work, we build on and extend the analysis presented in Ref.~\cite{AraujoFilho:2026zyt} (also based on the analysis of Ref. \cite{amarilo2024gravitational}), where gravitational wave propagation was examined in the presence of the CPT--even coefficient $\mathring{k}^{(4)}_{(I)}$ alone. Here, on the other hand, we generalize that framework by incorporating additional Lorentz-- and diffeomorphism--violating effects within the linearized gravitational sector of the Standard Model Extension. In this manner, focusing on isotropic contributions, we investigate the combined influence of the CPT--even dimension four coefficient $\mathring{k}^{(4)}_{(I)}$ and the CPT--odd dimension five coefficient $\mathring{k}^{(5)}_{(V)}$ on tensor gravitational waves. {
The modified propagation law leads to a rescaling of the propagation speed together with helicity--dependent corrections, giving rise to birefringence and polarization mixing without introducing extra propagating degrees of freedom.} We derive the corresponding retarded Green function and obtain explicit expressions for the gravitational waveforms generated by matter sources. As a concrete application, we consider a binary black hole system and demonstrate how Lorentz violation alters the observed strain through shifts in the retarded time, amplitude rescaling, and higher--derivative corrections to the quadrupole formula. Finally, using published LIGO--Virgo--KAGRA results, we translate current propagation and polarization constraints into bounds on $\mathring{k}^{(4)}_{(I)}$ and $\mathring{k}^{(5)}_{(V)}$.

%%%%%%%%%%%%%%%%%%%%%%%%%%%%%%%%%%%%%%%%%%%%%%%%%%%%%%%%%%%%%%%%%%%%%%%%%%%%%%%%%%%%%%%%%%%%%%%%%%%%%%%%%%%%%%%%%%%%%%%%%%%%%%%%%%%%%%%%%%%%%%%%%%%%%%%%%%%%%%%%%%%%%%%%%%%%%%%%%%%%%%%%%%%%%%%%%%%%%%%%%%%%%%%%%%%%%%%%%%%%%%%%%%%%%%%%%%%%%%%%%%%%%%%%%%%%%%%%%%%%%%%%%%%%%%%%%%%%%%%%%%%%%%%%%%%%%%%%%%%%%%%%%%%%%%%%%%%%%%%%%%%%%%%%%%%%%%%%%%%%%%%%%%%%%%%%%%%%%%%%%%%%%%%%%%%%%%

\section{Theoretical framework}

We begin by outlining the treatment of gravitational radiation within the linearized gravitational sector of the Standard Model Extension. In this setting, departures from exact Lorentz and diffeomorphism invariance are incorporated through additional operators that act directly on the tensor perturbations of the metric. These terms alter the propagation properties of gravitational waves, producing modified dispersion relations that deviate from the relativistic form. The formulation relies on the quadratic expansion of the gravitational action about flat spacetime, where the effects of symmetry breaking are encoded as corrections to the free graviton dynamics \cite{Kostelecky:2016kfm,Wang:2025fhw}
\bqn \lb{calL}
\mathcal{L} &=& \frac{1}{4}\epsilon^{\mu \rho \alpha \kappa} \epsilon^{\nu \sigma \beta \lambda} \eta_{\kappa \lambda} h_{\mu\nu} \partial_\alpha \partial_\beta h_{\rho \sigma} \nb + \frac{1}{4}h_{\mu\nu} \sum_{{\cal K}, d}  \hat{\cal K}^{(d)\mu\nu\rho\sigma}h_{\rho \sigma}.
\eqn
To describe weak gravitational fields, the spacetime metric is expanded about a flat background as $g_{\mu\nu}=\eta_{\mu\nu}+h_{\mu\nu}$, with $\eta_{\mu\nu}$ representing the Minkowski metric and $h_{\mu\nu}$ the dynamical perturbation. Within this setup, the symbol $\epsilon^{\mu\rho\alpha\kappa}$ denotes the fully antisymmetric Levi--Civita tensor. At quadratic order in $h_{\mu\nu}$, the action reduces to the familiar linearized Einstein–Hilbert form, governing the standard propagation of tensor modes. Deviations from this behavior arise through additional contributions that violate Lorentz and diffeomorphism symmetries. Such effects are incorporated by means of differential operators $\hat{\cal K}^{(d)\mu\nu\rho\sigma}$, which are built from constant background tensors contracted with spacetime derivatives and organized according to their mass dimension
${\cal K}^{(d)\mu\nu\rho\sigma\alpha_1\ldots\alpha_{d-2}}$ with $d-2$ spacetime derivatives $\partial_{\alpha_1}\cdots\partial_{\alpha_{d-2}}$
\bqn
\hat{\cal K}^{(d)\mu\nu\rho\sigma} = {\cal K}^{(d)\mu\nu\rho\sigma \alpha _1 \alpha _2 \cdots \alpha _{d-2}}{\partial_{\alpha _1 } \partial_{\alpha _2} \cdots \partial_{\alpha _{d-2}}}.
\eqn

The coefficients ${\cal K}^{(d)\mu\nu\rho\sigma\alpha_1\ldots\alpha_{d-2}}$ are assigned mass dimension $4-d$ and are assumed to represent small, fixed background quantities on the spacetime scales relevant for gravitational wave propagation. These tensors enter the linearized dynamics only through specific index structures of the associated differential operators $\hat{\cal K}^{(d)\mu\nu\rho\sigma}$. In particular, nontrivial contributions arise when the operator fails to exhibit symmetry under the interchange of the index pairs $(\mu\nu)$ and $(\rho\sigma)$. This condition can be expressed as $K^{(d)(\mu\nu)(\rho\sigma)} \neq \pm K^{(d)(\rho\sigma)(\mu\nu)}$, where the choice of the plus or minus sign is determined by whether the operator dimension $d$ is odd or even, respectively \cite{Kostelecky:2016kfm,Wang:2025fhw}
\bqn
\hat{\cal K}^{(d)\mu\nu\rho\sigma} = {\Tilde{\hat{s}}}^{(d)\mu\rho\nu\sigma}+{\Tilde{\hat{q}}}^{(d)\mu\rho\nu\sigma}+{\Tilde{\hat{k}}}^{(d)\mu\rho\nu\sigma}.
\eqn
Based on their index symmetries, Lorentz--violating operators appearing in the quadratic gravitational action can be grouped into a small number of distinct categories. This classification is determined by the transformation properties of their tensor indices under permutations. One subset consists of operators $\Tilde{\hat{s}}^{(d)\mu\rho\nu\sigma}$ that are antisymmetric within both index pairs $(\mu\rho)$ and $(\nu\sigma)$. A second subset, labeled $\Tilde{\hat{q}}^{(d)\mu\rho\nu\sigma}$, exhibits mixed symmetry, being antisymmetric in $(\mu\rho)$ while remaining symmetric in $(\nu\sigma)$. The final subset, denoted $\Tilde{\hat{k}}^{(d)\mu\rho\nu\sigma}$, is constructed to be completely symmetric under all index interchanges. Each of these operator families can be further resolved into irreducible tensor components, a step that clarifies the distinct ways in which they modify the propagation of gravitational waves. Carrying out this decomposition leads to a total of fourteen independent operator classes, as cataloged in Table~I of Refs.~\cite{Kostelecky:2016kfm,Wang:2025fhw}.

Within this classification, the $\Tilde{\hat{s}}$--sector plays a special role. The operators $\Tilde{\hat{s}}^{(d)\mu\rho\nu\sigma}$ are CPT--even, remaining unchanged under the combined action of charge conjugation, parity, and time reversal. Their tensorial structure allows them to be expressed in terms of three separate irreducible contributions, which can be analyzed independently in the context of gravitational wave dynamics
\bqn
\Tilde{\hat{s}}^{{(d)\mu\rho\nu\sigma}} = \Tilde{\hat{s}}^{(d, 0)\mu\rho\nu\sigma} + \Tilde{\hat{s}}^{(d, 1)\mu\rho\nu\sigma} +  \Tilde{\hat{s}}^{(d, 2)\mu\rho\nu\sigma},
\eqn
with we have
\bqn\lb{sd}
{\Tilde{\hat{s}}}^{(d,0)\mu \rho \nu \sigma} &=& {s}^{(d,0) \mu \rho \alpha_{1} \nu \sigma \alpha_{2} \alpha_3 \ldots \alpha_{d-2}} \partial_{\alpha_{1}} \ldots \partial_{\alpha_{d-2}}, \lb{sd0}\nb\\
{\Tilde{\hat{s}}}^{(d,1)\mu \rho \nu \sigma} &=& {s}^{(d,1) \mu \rho \nu \sigma \alpha_{1}  \ldots \alpha_{d-2}} \partial_{\alpha_{1}} \ldots \partial_{\alpha_{d-2}}, \lb{sd1}\nb\\
{\Tilde{\hat{s}}}^{(d,2)\mu \rho \nu \sigma} &=& {s}^{(d,2) \mu \rho \alpha_{1} \nu \sigma \alpha_{2} \alpha_3 \ldots \alpha_{d-2}} \partial_{\alpha_{1}} \ldots \partial_{\alpha_{d-2}}. \lb{sd2}
\eqn

A different behavior characterizes the operators belonging to the $\Tilde{\hat{q}}$ sector. The tensors $\Tilde{\hat{q}}^{(d)\mu\rho\nu\sigma}$ are CPT--odd and consequently change sign under the simultaneous application of charge conjugation, parity, and time reversal. Owing to their mixed symmetry properties, these operators admit a decomposition into six mutually independent irreducible components, each contributing in a distinct manner to modifications of gravitational wave propagation
\bqn
\Tilde{\hat{q}}^{(d)\mu\rho\nu\sigma} &=& \Tilde{\hat{q}}^{(d, 0)\mu\rho\nu\sigma}+\Tilde{\hat{q}}^{(d, 1)\mu\rho\nu\sigma}+\Tilde{\hat{q}}^{(d, 2)\mu\rho\nu\sigma}\nb +\Tilde{\hat{q}}^{(d, 3)\mu\rho\nu\sigma}+\Tilde{\hat{q}}^{(d, 4)\mu\rho\nu\sigma}+\Tilde{\hat{q}}^{(d, 5)\mu\rho\nu\sigma},\nb\\
\eqn
where
\bqn\lb{qd}
\Tilde{\hat{q}}^{(d, 0)\mu\rho\nu\sigma}&=&\Tilde{\hat{q}}^{(d, 0)\mu\rho\alpha_1 \nu \alpha_2 \sigma \alpha_3 \alpha_4 \ldots \alpha_{d-2}} \partial_{\alpha_{1}} \ldots \partial_{\alpha_{d-2}}, \nb\\
\Tilde{\hat{q}}^{(d, 1)\mu\rho\nu\sigma}&=&\Tilde{\hat{q}}^{(d, 1)\mu\rho \nu  \sigma \alpha_1 \alpha_2 \ldots \alpha_{d-2}} \partial_{\alpha_{1}} \ldots \partial_{\alpha_{d-2}}, \nb\\
\Tilde{\hat{q}}^{(d, 2)\mu\rho\nu\sigma}&=&\Tilde{\hat{q}}^{(d, 2)\mu\rho \nu \alpha_1 \sigma \alpha_2  \ldots \alpha_{d-2}} \partial_{\alpha_{1}} \ldots \partial_{\alpha_{d-2}}, \nb\\
\Tilde{\hat{q}}^{(d, 3)\mu\rho\nu\sigma}&=&\Tilde{\hat{q}}^{(d, 3)\mu\rho\alpha_1 \nu  \sigma \alpha_2 \ldots \alpha_{d-2}} \partial_{\alpha_{1}} \ldots \partial_{\alpha_{d-2}}, \nb\\
\Tilde{\hat{q}}^{(d, 4)\mu\rho\nu\sigma}&=&\Tilde{\hat{q}}^{(d, 4)\mu\rho\nu \alpha_1 \sigma \alpha_2 \alpha_3 \alpha_4 \ldots \alpha_{d-2}} \partial_{\alpha_{1}} \ldots \partial_{\alpha_{d-2}}, \nb\\
\Tilde{\hat{q}}^{(d, 5)\mu\rho\nu\sigma}&=&\Tilde{\hat{q}}^{(d, 5)\mu\rho \alpha_1 \nu \alpha_2 \sigma \alpha_3 \alpha_4 \ldots \alpha_{d-2}} \partial_{\alpha_{1}} \ldots \partial_{\alpha_{d-2}}.\nb\\
\eqn
It is worthy commenting that operators in the $\Tilde{\hat{k}}$ sector are CPT--even and can be resolved into five independent irreducible components
\bqn
\Tilde{\hat{k}}^{(d)\mu\nu\rho\sigma} &=& \Tilde{\hat{k}}^{(d, 0)\mu\nu\rho\sigma} +\Tilde{\hat{k}}^{(d, 1)\mu\nu\rho\sigma} +\Tilde{\hat{k}}^{(d, 2)\mu\nu\rho\sigma} \nb +\Tilde{\hat{k}}^{(d, 3)\mu\nu\rho\sigma} +\Tilde{\hat{k}}^{(d, 4)\mu\nu\rho\sigma} ,
\eqn
in a such way that
\bqn\lb{kd}
\Tilde{\hat{k}}^{(d, 0)\mu\nu\rho\sigma} &=& \Tilde{\hat{k}}^{(d, 0)\mu \alpha_1 \nu \alpha_2 \rho \alpha_3 \sigma \alpha_4 \alpha_5 \ldots \alpha_{d-2}}  \partial_{\alpha_{1}} \ldots \partial_{\alpha_{d-2}}, \nb\\
\Tilde{\hat{k}}^{(d, 1)\mu\nu\rho\sigma} &=& \Tilde{\hat{k}}^{(d, 1)\mu \nu  \rho \sigma \alpha_1 \ldots \alpha_{d-2}}  \partial_{\alpha_{1}} \ldots \partial_{\alpha_{d-2}}, \nb\\
\Tilde{\hat{k}}^{(d, 2)\mu\nu\rho\sigma} &=& \Tilde{\hat{k}}^{(d, 2)\mu \alpha_1 \nu  \rho  \sigma \alpha_1 \alpha_2 \ldots \alpha_{d-2}}  \partial_{\alpha_{1}} \ldots \partial_{\alpha_{d-2}}, \nb\\
\Tilde{\hat{k}}^{(d, 3)\mu\nu\rho\sigma} &=& \Tilde{\hat{k}}^{(d, 3)\mu \alpha_1 \nu \alpha_2 \rho  \sigma \alpha_3 \alpha_5 \ldots \alpha_{d-2}}  \partial_{\alpha_{1}} \ldots \partial_{\alpha_{d-2}}, \nb\\
\Tilde{\hat{k}}^{(d, 4)\mu\nu\rho\sigma} &=& \Tilde{\hat{k}}^{(d, 4)\mu \alpha_1 \nu \alpha_2 \rho \alpha_3 \sigma \alpha_4 \alpha_5 \ldots \alpha_{d-2}}  \partial_{\alpha_{1}} \ldots \partial_{\alpha_{d-2}}. \nb\\
\eqn
At the level of linearized gravitational dynamics, all possible Lorentz--violating effects can be parametrized by a finite set of coefficients. This construction leads to fourteen independent coefficient classes, whose defining symmetry properties and tensorial structures are summarized in Table~I of Refs.~\cite{Kostelecky:2016kfm,Wang:2025fhw}, as argued before. Taken together, these coefficients exhaust the space of allowed modifications to graviton propagation at quadratic order and capture every observable deviation from the standard relativistic behavior of gravitational waves within this framework.

Requiring invariance under infinitesimal coordinate redefinitions imposes severe constraints on the form of these terms in the linearized theory. In particular, demanding that the quadratic action $S \sim \int \mathrm{d}^4x\,\mathcal{L}$ remain invariant under the gauge transformation $h_{\mu\nu} \rightarrow h_{\mu\nu} + \partial_\mu \xi_\nu + \partial_\nu \xi_\mu$ leads to a nontrivial condition on the allowed operator structure. The Lorentz--violating operator must then obey the relation $\hat{\cal K}^{(d)(\mu\nu)(\rho\sigma)}\partial_\nu = \pm \hat{\cal K}^{(d)(\rho\sigma)(\mu\nu)}\partial_\nu$, where the sign depends on the operator dimension.

Enforcing this constraint drastically reduces the space of admissible operators. The general Lorentz--violating sector collapses into three distinct classes, conventionally labeled as ${\Tilde{\hat{s}}}^{(d,0)\mu\rho\nu\sigma}$, ${\Tilde{\hat{q}}}^{(d,0)\mu\rho\nu\sigma}$, and ${\Tilde{\hat{k}}}^{(d,0)\mu\rho\nu\sigma}$ \cite{Kostelecky:2016kfm,Wang:2025fhw}. An additional hierarchy follows from symmetry considerations: if linearized diffeomorphism invariance is preserved, Lorentz--violating effects are restricted to operators with mass dimension $d \ge 4$, whereas relaxing this requirement allows contributions already at $d \ge 2$.

The equations that determine the propagation of gravitational waves are obtained by extremizing the quadratic action $S \sim \int \mathrm{d}^4x\,\mathcal{L}$ with respect to the metric fluctuation $h_{\mu\nu}$. Substituting the Lagrangian density introduced in Eq.~(\ref{calL}) and carrying out the functional variation yields
\bqn\lb{eom0}
\frac{1}{2}\eta_{\rho \sigma} \epsilon^{\mu\rho\alpha \kappa} \epsilon^{\nu\sigma \beta \lambda} \partial_\alpha \partial_\beta h_{\kappa \lambda} - \delta M^{\mu\nu \rho\sigma} h_{\rho \sigma}=0,
\eqn
where the tensor operators take the form
\bqn
\delta M^{\mu\nu\rho \sigma} &=& - \frac{1}{4} \left(\Tilde{\hat{s}}^{\mu\rho \nu \sigma} + \Tilde{\hat{s}}^{\mu \sigma \nu \rho}\right) - \frac{1}{2} \Tilde{\hat{k}}^{\mu \nu \rho \sigma} -\frac{1}{8} \left(\Tilde{\hat{q}}^{\mu \rho \nu \sigma} + \Tilde{\hat{q}}^{\nu \rho \mu \sigma} +\Tilde{\hat{q}}^{\mu \sigma \nu \rho} + \Tilde{\hat{q}}^{\nu \sigma \mu \rho}\right).\nb\\
\eqn
so that
\bqn
\Tilde{\hat{s}}^{\mu\rho \nu \sigma}  = \sum_{d} \Tilde{\hat{s}}^{(d)\mu\rho \nu \sigma}, \\
\Tilde{\hat{q}}^{\mu\rho \nu \sigma}  = \sum_{d} \Tilde{\hat{q}}^{(d)\mu\rho \nu \sigma}, \\
\Tilde{\hat{k}}^{\mu\rho \nu \sigma}  = \sum_{d} \Tilde{\hat{k}}^{(d)\mu\rho \nu \sigma}.
\eqn

In the framework of General Relativity, gravitational waves propagate through a highly constrained channel, being described solely by two transverse traceless tensor polarizations. This economy of degrees of freedom relies on the exact validity of Lorentz invariance and diffeomorphism symmetry. Once these symmetries are relaxed, the polarization content of metric perturbations need not remain limited to the tensor sector. In more general settings, additional scalar and vector excitations may become dynamical, extending the spectrum of gravitational wave polarizations beyond the standard picture. In linearized gravity with Lorentz--violating terms however, such non--tensorial modes do not necessarily arise as independent sources. Instead, they can be activated through couplings to the tensor perturbations themselves, leading to indirect excitation mechanisms, as discussed in Refs.~\cite{Hou:2024xbv,Wang:2025fhw}. In other words, the presence and relevance of these extra components depend sensitively on the structure of the symmetry--breaking operators.

From an observational standpoint, however, no deviations from the tensor--only polarization pattern have been detected so far. Analyses performed by the LIGO--Virgo--KAGRA Collaboration are fully compatible with the propagation of two tensor modes and show no statistically significant evidence for scalar or vector polarizations \cite{KAGRA:2021vkt}. Moreover, even in scenarios where such modes exist, their contributions are expected to be strongly suppressed relative to the dominant tensor signal.

Guided by these experimental findings and aiming for a focused theoretical treatment, we restrict attention to the transverse traceless tensor sector in what follows. Any additional polarizations are assumed to be negligible, and the analysis proceeds in close analogy with Ref.~\cite{Kostelecky:2016kfm}, concentrating on how Lorentz-- and diffeomorphism--violating operators modify the propagation of the two tensorial gravitational wave modes alone.

%%%%%%%%%%%%%%%%%%%%%%%%%%%%%%%%%%%%%%%%%%%%%%%%%%%%%%%%%%%%%%%%%%%%%%%%%%%%%%%%%%%%%%%%%%%%%%%%%%%%%%%%%%%%%%%%%%%%%%%%%%%%%%%%%%%%%%%%%%%%%%%%%%%%%%%%%%%%%%%%%%%%%%%%%%%%%%%%%%%%%%%%%%%%%%%%%%%%%%%%%%%%%%%%%%%%%%%%%%%%%%%%%%%%%%%%%%%%%%%%%%%%%%%%%%%%%%%%%%%%%%%%%%%%%%%%%%%%%%%%%%%%%%%%%%%%%%%%%%%%%%%%%%%%%%%%%%%%%%%%%%%%%%%%%%%%%%%%%%%%%%%%%%%%%%%%%%%%%%

\section{Polarization structure of Lorentz--violating gravitational waves}

For Lorentz--violating operators {in the linearized gravitational sector of the SME}, the modified dispersion relation {can be written} in the compact form \cite{Kostelecky:2016kfm}
\begin{equation}
\label{mdr5}
p^0=\Big(1-\varsigma_0 \pm \sqrt{\varsigma_1^2+\varsigma_2^2+\varsigma_3^2}\Big)\,|\vec p| .
\end{equation}
In this expression, $p^\mu=(p^0,\vec p)$ represents the four--momentum of the gravitational wave. The quantities $\varsigma_i$ collect all contributions associated with Lorentz violation and are constructed as momentum--dependent combinations of gauge--invariant operators entering the quadratic action for metric fluctuations. The detailed form of these functions can be found in Ref.~\cite{Kostelecky:2016kfm}
\begin{align}
\varsigma_0 &=
\frac{1}{4p^2}
\left\{
-\,\Tilde{\hat{s}}^{\mu\nu}{}_{\mu\nu}
+\frac{1}{2}\,\Tilde{\hat{k}}^{\mu\nu}{}_{\mu\nu}
\right\}, \\[1ex]
(\varsigma_1)^2+(\varsigma_2)^2 &=
\frac{1}{8p^4}
\left\{
\Tilde{\hat{k}}^{\mu\nu\rho\sigma}\Tilde{\hat{k}}_{\mu\nu\rho\sigma}
-\Tilde{\hat{k}}^{\mu\rho}{}_{\nu\rho}\,\Tilde{\hat{k}}_{\mu\sigma}{}^{\nu\sigma}
+\frac{1}{8}\,\Tilde{\hat{k}}^{\mu\nu}{}_{\mu\nu}\,
\Tilde{\hat{k}}^{\rho\sigma}{}_{\rho\sigma}
\right\}, \\[1ex]
(\varsigma_3)^2 &=
\frac{1}{16p^4}
\left\{
-\frac{1}{2}\,\Tilde{\hat{q}}^{\mu\rho\nu\sigma}\Tilde{\hat{q}}_{\mu\rho\nu\sigma}
-\Tilde{\hat{q}}^{\mu\nu\rho\sigma}\Tilde{\hat{q}}_{\mu\rho\nu\sigma}
+\big(\Tilde{\hat{q}}^{\mu\rho\nu}{}_{\rho}+\Tilde{\hat{q}}^{\nu\rho\mu}{}_{\rho}\big)
\Tilde{\hat{q}}_{\mu\sigma\nu}{}^{\sigma}
\right\}.
\end{align}
In momentum space, the Lorentz--violating operators are indicated by a hat and are obtained by substituting spacetime derivatives with factors of the wave four--momentum according to $\partial_\mu \to i p_\mu$.

The modified dispersion relation in Eq.~\eqref{mdr5} shows that Lorentz violation can affect gravitational wave propagation in three fundamentally distinct ways. One class of effects alters the angular response of the waves. If rotational symmetry is not preserved, the propagation becomes anisotropic, and the resulting behavior depends---within a fixed observer frame---on coefficients carrying explicit spatial indices. A separate modification concerns the dependence of the propagation speed on frequency. Such dispersion emerges whenever the effective operators depend on the wave momentum. In contrast, frequency--independent propagation is possible only in the special case of mass dimension $d=4$ operators, where the dynamics are controlled solely by the tensor $\Tilde{\hat{s}}^{\mu\rho\nu\sigma}$.
Finally, Lorentz violation may induce a separation of polarization states during propagation. The existence of two distinct branches in the solutions of Eq.~\eqref{mdr5} reflects birefringence, signaling that different polarizations travel with different phase velocities. This phenomenon arises from the $\Tilde{\hat{q}}^{\mu\rho\nu\sigma}$ and $\Tilde{\hat{k}}^{\mu\nu\rho\sigma}$ sectors and therefore requires operators of mass dimension $d>4$. {In the isotropic truncation adopted below, only the birefringent contribution associated with $\mathring{k}^{(5)}_{(V)}$ is retained, while directional anisotropies are not included.}

{For isotropic coefficients,} the dispersion relation can be written in the series form \cite{Kostelecky:2016kfm}
\begin{equation}
\label{kkkk}
\omega
=
\left(1-\mathring{k}^{(4)}_{(I)}\right)\,|\mathbf{p}|
\;\pm\;
\mathring{k}^{(5)}_{(V)}\,\omega^{2}
\;-\;
\mathring{k}^{(6)}_{(I)}\,\omega^{3}
\;\pm\;
\mathring{k}^{(7)}_{(V)}\,\omega^{4}
\;-\;
\mathring{k}^{(8)}_{(I)}\,\omega^{5}
\;\pm\;\ldots
\end{equation}

Recently, Ref.~\cite{AraujoFilho:2026zyt} restricted its analysis to the case $d=4$, retaining only the coefficient
$\mathring{k}^{(4)}_{(I)}$. In that paper, no birefringence {or} anisotropies were reported. {In the present work, we keep this $d=4$ contribution and include the leading CPT--odd higher--dimension correction controlled by $\mathring{k}^{(5)}_{(V)}$. Since both coefficients are treated in their isotropic sector, the setup considered here does not generate directional anisotropies; instead, the new physical effect is helicity--dependent propagation, which produces birefringence and polarization mixing.}

The polarization states carried by gravitational waves provide direct access to both the properties of their sources and the symmetry principles underlying the gravitational interaction. When local Lorentz symmetry is not exact, its violation can leave characteristic signatures in the polarization sector, altering how different modes propagate and mix. As a result, measurements of gravitational wave polarization offer a sensitive observational window into both modified gravitational dynamics and nonstandard propagation effects \cite{le2017theory,liang2017polarizations,zhang2018velocity,hou2018polarizations,mewes2019signals,liang2022polarizations}.

We work within a weak--field regime in which gravitational waves propagate over a Minkowski background characterized by the metric $\eta_{\mu\nu}=\mathrm{diag}(-1,1,1,1)$. Focusing exclusively on the tensorial degrees of freedom, the spatial components of the metric fluctuation, $h_{ij}(x)$ with indices $i,j=1,2,3$, satisfy a wave equation whose form is altered by Lorentz--violating contributions arising in the linearized theory
\begin{equation}
\label{EOM_SME_TT_new}
\big(\partial_t^2-\nabla^2\big)\,h_{ij}(x)
\;+\;2\,\delta M_{ijmn}(\partial)\,h_{mn}(x)=0 .
\end{equation}
All derivatives are defined with respect to the flat background geometry, so that $\partial_t$ denotes differentiation with respect to the coordinate time $t$ and the spatial Laplacian is given by $\nabla^2=\delta^{kl}\partial_k\partial_l$. Deviations from standard wave propagation are captured by the operator $\delta M_{ijmn}(\partial)$, which represents the Lorentz--violating corrections induced by higher--mass--dimension terms in the gravitational sector. Acting linearly on the tensor perturbations, this operator can include both CPT--even and CPT--odd contributions. Although these terms modify the dispersion and polarization behavior of gravitational waves, they do so without affecting the underlying Minkowski background structure.

{In the tensor--sector approximation adopted here, we impose} the transverse traceless conditions
\begin{equation}
\label{constraints_new}
h_{ii}=0,
\qquad
\partial_i h_{ij}=0 .
\end{equation}
{Within this projected sector,} no independent scalar or vector modes propagate, and the physical content of the theory remains restricted to two tensorial degrees of freedom. The Lorentz--violating terms introduced in this analysis therefore do not enlarge the polarization basis. Instead, their influence is encoded in changes to the dispersion relations and in modifications to the way the existing tensor modes propagate through spacetime.

With this setup, the polarization structure can be examined by expressing the metric perturbation as a superposition of plane wave modes of the form
\begin{equation}
h_{\mu\nu}(x)=\varepsilon_{\mu\nu}\,e^{i p_\mu x^\mu},
\end{equation}
with
\begin{equation}
p^\mu=(\omega,0,0,p),
\end{equation}
describing a wave traveling in the $z$ direction. Imposing transversality with respect to the wave four--momentum, we obtain
\begin{equation}
p^\mu \varepsilon_{\mu\nu}=0 .
\end{equation}
Notice that, when combined with the tracelessness condition $\varepsilon^\mu{}_\mu=0$, these constraints restrict the polarization tensor to components that lie entirely within the spatial subspace and are orthogonal to the direction of propagation. As a result, only transverse spatial elements survive. Adopting the conventional linear polarization basis, the tensor structure can then be written as
\begin{equation}
\varepsilon_{\mu\nu}=
\begin{pmatrix}
0 & 0 & 0 & 0 \\
0 & \varepsilon_{11} & \varepsilon_{12} & 0 \\
0 & \varepsilon_{12} & -\varepsilon_{11} & 0 \\
0 & 0 & 0 & 0
\end{pmatrix}.
\end{equation}
The quantities $\varepsilon_{11}$ and $\varepsilon_{12}$ parametrize the two independent degrees of freedom associated with the transverse traceless tensor sector. When CPT--odd Lorentz--violating operators with mass dimension $d>4$ are present, {the linear $(+,\times)$ basis no longer diagonalizes the propagation}. Instead, a relative phase develops {between the two helicity states}, causing the polarization state to evolve dynamically. This effect manifests as a continuous rotation within the $(+,\times)$ polarization basis as the gravitational wave travels through spacetime.

When Lorentz--violating operators of mass dimension $d=5$ are present, in particular those controlled by the CPT--odd coefficient $\mathring{k}^{(5)}_{(V)}$, the structure of the polarization tensor itself remains transverse and traceless, but the physical propagation eigenstates are no longer given by the linear $(+,\times)$ modes. In other words, the modified dispersion relation,
\begin{equation}
\omega
=
\left(1-\mathring{k}^{(4)}_{(I)}\right)\,|\mathbf{p}|
\;\pm\;
\mathring{k}^{(5)}_{(V)}\,\omega^{2},
\end{equation}
indicates that the two helicity states propagate differently. To leading order in Lorentz violation, this relation yields
\begin{equation}
\omega_{\lambda}\simeq
v|\mathbf{p}|
+
\lambda\,\mathring{k}^{(5)}_{(V)}\,v^{2}|\mathbf{p}|^{2}
{\,=\,}
\left(1-\mathring{k}^{(4)}_{(I)}\right)|\mathbf{p}|
+
\lambda\,\mathring{k}^{(5)}_{(V)}\,|\mathbf{p}|^{2}
{\,+\,\mathcal{O}\!\left(\mathring{k}^{(4)}_{(I)}\mathring{k}^{(5)}_{(V)}\right)},
\qquad
\lambda=\pm1 ,
\end{equation}
where {$v\equiv1-\mathring{k}^{(4)}_{(I)}$ and} $\lambda$ labels the right-- and left--handed circular polarizations. {In the first equality, the factor $v^{2}$ in the CPT--odd correction follows directly from solving the dispersion relation perturbatively. In the second equality, mixed contributions of order $\mathring{k}^{(4)}_{(I)}\mathring{k}^{(5)}_{(V)}$ have been neglected consistently with the first--order treatment used below.}

It is therefore convenient to introduce the circular polarization basis,
\begin{equation}
h_{R,L}=\frac{1}{\sqrt{2}}\big(h_+ \mp i\,h_\times\big),
\end{equation}
which diagonalizes the propagation in the presence of the CPT--odd term. The two helicity modes acquire different phase velocities and {, equivalently, at fixed observed angular frequency, different wavenumbers,} accumulating a relative phase during propagation. After traveling a distance $L$, this phase difference is given by
\begin{equation}
\Delta\phi
=
{\big[p_{-}(\Omega)-p_{+}(\Omega)\big]\,L}
\simeq
{\frac{2\,\mathring{k}^{(5)}_{(V)}}{v}\,\Omega^{2}L} ,
\end{equation}
{where the sign convention follows from the definition adopted for the circular polarization basis. Here $\Omega$ denotes the gravitational wave angular frequency. In the binary application discussed later, $\Omega=2\omega$.}
Transforming back to the linear polarization basis, the plus and cross amplitudes become mixed according to
\begin{equation}
\begin{pmatrix}
h_+(L)\\
h_\times(L)
\end{pmatrix}
=
\begin{pmatrix}
\cos(\Delta\phi/2) & -\sin(\Delta\phi/2)\\
\sin(\Delta\phi/2) & \cos(\Delta\phi/2)
\end{pmatrix}
\begin{pmatrix}
h_+(0)\\
h_\times(0)
\end{pmatrix}.
\end{equation}

This result shows that, although no additional tensor degrees of freedom are introduced {within the TT--projected sector considered here}, the presence of the $d=5$ CPT--odd operator leads to helicity--dependent propagation and induces a rotation of the polarization basis. As a consequence, an initially linearly polarized { component of a} gravitational wave remains linearly polarized, but with a rotated polarization angle. In other words, this phenomenon corresponds to gravitational wave birefringence and provides a {direct} observational signature of higher--dimension Lorentz--violating operators in the gravitational sector, as we shall {discuss} in the forthcoming sections.

%%%%%%%%%%%%%%%%%%%%%%%%%%%%%%%%%%%%%%%%%%%%%%%%%%%%%%%%%%%%%%%%%%%%%%%%%%%%%%%%%%%%%%%%%%%%%%%%%%%%%%%%%%%%%%%%%%%%%%%%%%%%%%%%%%%%%%%%%%%%%%%%%%%%%%%%%%%%%%%%%%%%%%%%%%%%%%%%%%%%%%%%%%%%%%%%%%%%%%%%%%%%%%%%%%%%%%%%%%%%%%%%%%%%%%%%%%%%%%%%%%%%%%%%%%%%%%%%%%%%%%%%%%%%%%%%%%%%%%%%%%%%%%%%%%%%%%%%%%%%%%%%%%%%%%%%%%%%%%%%%%%%%%%%%%%%%%%%%%%%%%%%%%%%%%%%%%%%%%%%%%%%%%%%%%%%%%%%%%%%%%%%%%%%%%%%%%%%%%%%%%%%%%%%%%%%%%%%%%%%%%%%%%%%%%%%%%%%%%%%%%%%%%%%%%%%%%%

\section{Gravitational wave emission from matter sources}

To account for gravitational wave generation, the homogeneous equations governing metric perturbations are extended by coupling them to an external tensor source $J_{ij}$. Introducing this current into the linearized framework defined by Eq.~\eqref{EOM_SME_TT_new} converts the vacuum propagation equation into a sourced system, yielding the modified field equation
\begin{equation}
  \big(\partial_t^2-\nabla^2\big)\,h_{ij}(x)
\;+\;2\,\delta M_{ijmn}(\partial)\,h_{mn}(x)
=
J_{ij}(x).
\end{equation}

The solution for the metric fluctuation follows by convolving the source with the retarded Green function associated with the modified wave operator, leading to
\begin{equation}
h_{ij}(x)
=
\sum_{\lambda=\pm}
\int \mathrm{d}^4y \;
G^{\mathrm{ret}}_{\lambda}(x-y)\,
\Pi^{(\lambda)}_{ijmn}\,
J_{mn}(y),
\end{equation}
where $G^{\mathrm{ret}}_{\lambda}(x-y)$ denotes the retarded Green function associated with the helicity eigenmode $\lambda=\pm$, and $\Pi^{(\lambda)}_{ijmn}$ is the transverse--traceless projector onto the corresponding circular polarization subspace. This decomposition makes explicit that, in the presence of CPT--odd Lorentz--violating operators, the two tensor helicities propagate independently with distinct dispersion relations, while the physical source entering the field equation is the transverse--traceless projection of the stress--energy tensor.

The construction of $G^{\mathrm{ret}}(x-y)$ is guided by the modified dispersion relation introduced in Eq.~\eqref{kkkk}. Restricting to the isotropic Lorentz--violating coefficients $\mathring{k}^{(4)}_{(I)}$ and $\mathring{k}^{(5)}_{(V)}$, the dispersion relation reads
\begin{equation}
\label{modddss}
\omega = p^{0}
=
v\,|\vec{p}|
\;\pm\;
\mathring{k}^{(5)}_{(V)}\,\omega^{2},
\qquad
v \equiv 1 - \mathring{k}^{(4)}_{(I)} ,
\end{equation}
where the $\pm$ sign labels the two circular polarization (helicity) states. In this isotropic truncation, Lorentz violation modifies the propagation speed and induces birefringence, but does not introduce directional anisotropies.

In Fourier space, the retarded Green function associated with each helicity mode is therefore {represented, up to first order in $\mathring{k}^{(5)}_{(V)}$, by}
\begin{equation}
\tilde{G}^{\mathrm{ret}}_{\pm}(p)
=
\frac{1}
{v^{2} |\vec{p}|^{2}-(p^{0})^{2} 
\;\pm\;
2\,\mathring{k}^{(5)}_{(V)}\, (p^{0})^{3}
- i\epsilon p^{0}} \, ,
\end{equation}
where the $i\epsilon p^{0}$ prescription enforces causal propagation.

Assuming $|\mathring{k}^{(5)}_{(V)}| \ll 1$, the Green function can be expanded perturbatively to first order as
\begin{equation}
\tilde{G}^{\mathrm{ret}}_{\pm}(p)
\simeq
\frac{1}{v^{2} |\vec{p}|^{2} -(p^{0})^{2}  - i\epsilon p^{0}}
\;\mp\;
2\,\mathring{k}^{(5)}_{(V)}
\frac{(p^{0})^{3}}
{\big[v^{2} |\vec{p}|^{2} -(p^{0})^{2} - i\epsilon p^{0}\big]^{2}} .
\label{perk5}
\end{equation}
The inverse Fourier transform then yields
\begin{equation}
G^{\mathrm{ret}}_{\pm}(t,\vec{x})
=
\int \frac{\mathrm{d}p^{0}}{2\pi}
\int \frac{\mathrm{d}^{3}p}{(2\pi)^{3}}
\,
e^{-ip^{0}t + i\vec{p}\cdot\vec{x}}\,
\tilde{G}^{\mathrm{ret}}_{\pm}(p) .
\label{sdfffff}
\end{equation}

{To make the inverse Fourier transform explicit, we first evaluate the frequency integral and then perform the remaining radial momentum integral. After the $p^{0}$ integration, the zeroth--order (the first term after the symbol $\simeq$ in Eq. (\ref{perk5})) contribution gives
\begin{equation}
\widetilde{G}_{0}(t,k)
=
\Theta(t)\frac{\sin(vkt)}{vk},
\end{equation}
while the first--order contribution (the second term after the symbol $\simeq$ in Eq. (\ref{perk5})) in the CPT--odd coefficient gives
\begin{equation}
\widetilde{G}_{1}^{(\lambda)}(t,k)
=
i\,\mathring{k}^{(5)}_{(V)}\lambda\,\Theta(t)
\left[
2\cos(vkt)-vkt\sin(vkt)
\right].
\end{equation} 
The coordinate space Green function represented in Eq. (\ref{sdfffff}) follows from the radial inverse transform
\begin{equation}
G^{\rm ret}_{\lambda}(t,r)
=
\frac{1}{2\pi^{2}r}
\int_{0}^{\infty}
\mathrm{d}k\,k\sin(kr)
\left[
\widetilde{G}_{0}(t,k)
+
\widetilde{G}_{1}^{(\lambda)}(t,k)
\right].
\end{equation}
The leading term is evaluated through the distributional identity
\begin{equation}
\int_{0}^{\infty}\mathrm{d}k\,
\sin(kr)\sin(vkt)
=
\frac{\pi}{2}
\bigg[
\delta(r-vt)-\delta(r+vt)
\bigg].
\end{equation}
On the other hand, for the first--order CPT--odd sector, the required identities are
\begin{equation}
\int_{0}^{\infty}\mathrm{d}k\,
k\sin(kr)\cos(vkt)
=
-\frac{\pi}{2}
\bigg[
\delta'(r-vt)+\delta'(r+vt)
\bigg],
\end{equation}
and
\begin{equation}
\int_{0}^{\infty}\mathrm{d}k\,
k^{2}\sin(kr)\sin(vkt)
=
\frac{\pi}{2}
\bigg[
\delta''(r+vt)-\delta''(r-vt)
\bigg].
\end{equation}
Retaining only the physical retarded support, namely the contribution localized on $r=vt$ for $r>0$ and $t>0$, we obtain
\begin{equation}
G^{\rm ret}_{\lambda}(t,r)
=
\Theta(t)\frac{\delta(r-vt)}{4\pi v r}
+
\frac{i\,\mathring{k}^{(5)}_{(V)}\lambda\,\Theta(t)}{2\pi r}
\left[
-\delta'(r-vt)
+
\frac{vt}{2}\delta''(r-vt)
\right].
\label{sdsadasdas}
\end{equation}
Notice that the first term on the right--hand side reproduces the expression recently obtained in the literature by considering only $\mathring{k}^{(4)}_{(I)}$ \cite{AraujoFilho:2026zyt}. The second term, in turn, arises due to the contribution associated with $\mathring{k}^{(5)}_{(V)}$.} { At first sight, we might expect the imaginary contribution induced by $\mathring{k}^{(5)}_{(V)}$ in Eq.~(\ref{sdsadasdas}) could introduce difficulties in extracting physically meaningful quantities. However, as shown below, although this term appears explicitly in the helicity--resolved retarded Green function, it does not lead to an unphysical waveform. Once the strain is recombined and written in the real linear polarization basis, the apparent imaginary contribution is removed, yielding real observable components. Furthermore, in other words, the above expression} makes explicit the distinct roles of the Lorentz--violating coefficients. The coefficient $\mathring{k}^{(4)}_{(I)}$ enters through the rescaling of the propagation speed $v$, leading to a shift in the retarded time, while the CPT--odd coefficient $\mathring{k}^{(5)}_{(V)}$ induces helicity--dependent corrections encoded in higher derivatives of the light--cone delta distribution.

{In the weak field regime of general relativity, the observable gravitational wave signal in the radiation zone is carried by the spatial components of the metric perturbation. At leading order, the standard quadrupole formula gives
\begin{equation}
h_{ij}^{\rm GR}(t,{\bf r})
=
\frac{2G}{r}\,
\frac{\mathrm{d}^{2} I_{ij}}{\mathrm{d}t^{2}}(t-r),
\end{equation}
where
\begin{equation}
I_{ij}(t)
=
\int y^{i} y^{j} T^{00}(t,\vec y)\,\mathrm{d}^{3}y
\end{equation}
is the mass quadrupole moment of the source.

Once Lorentz--violating contributions are introduced through the dispersion relation in Eq.~\eqref{modddss}, the retarded propagation kernel is modified. In the helicity--resolved sector, the physical branch of the retarded Green function can be written, up to first order in $\mathring{k}^{(5)}_{(V)}$, as shown in Eq. (\ref{sdsadasdas}). Moreover, to obtain the waveform, we perform the convolution of the Green function with the quadrupole source. In the radiation zone, we define
\begin{equation}
F_{ij}(t')\equiv \frac{1}{2}\frac{\mathrm{d}^{2}I_{ij}}{\mathrm{d}{t'}^{2}}(t'),
\end{equation}
so that the helicity--resolved metric perturbation is obtained from
\begin{equation}
h_{ij}^{(\lambda)}(t,{\bf r})
=
16\pi G
\int_{-\infty}^{+\infty}
\mathrm{d}t'\,
G^{\rm ret}_{\lambda}(t-t',r)
F_{ij}(t') .
\end{equation}
The relevant distributional identity is
\begin{equation}
\int_{-\infty}^{+\infty}
\mathrm{d}t'\,
\delta^{(n)}
\left[
r-v(t-t')
\right]
H(t')
=
\frac{(-1)^n}{v^{n+1}}
\frac{\mathrm{d}^{n}H}{\mathrm{d}{t'}^{n}}
\bigg|_{t'=t_r},
\end{equation}
where $t_r=t-r/v$. For the leading term, this gives
\begin{equation}
\int_{-\infty}^{+\infty}
\mathrm{d}t'\,
\delta\left[
r-v(t-t')
\right]
F_{ij}(t')
=
\frac{1}{v}F_{ij}(t_r)
=
\frac{1}{2v}
I_{ij}^{(2)}(t_r).
\end{equation}
Thus the zeroth--order convolution with the modified propagation speed yields
\begin{equation}
16\pi G
\int_{-\infty}^{+\infty}
\mathrm{d}t'\,
\frac{\delta\left[r-v(t-t')\right]}{4\pi v r}
F_{ij}(t')
=
\frac{2G}{r}
\frac{I_{ij}^{(2)}(t_r)}{v^2}.
\end{equation}
This term corresponds to the usual quadrupole waveform evaluated at the shifted retarded time, together with the amplitude rescaling produced by the normalization of the modified Green function \cite{AraujoFilho:2026zyt}. In addition, this result reproduces Eq.~(40) of the recent paper \cite{AraujoFilho:2026zyt} up to an overall factor $1/v^{2}$. In that work, this factor was absorbed into the gravitational coupling. We note, however, a typographical point: the effective constant should read $\Tilde{G} \equiv G/v^{2}$, instead of simply $G$. In the present paper, we adopt a different convention by keeping the factor $1/v^{2}$ explicitly.

The correction controlled by $\mathring{k}^{(5)}_{(V)}$ comes from the derivative terms in Eq.~\eqref{sdsadasdas}. The two pieces must be convolved together:
\begin{equation}
\mathcal{C}_{ij}
\equiv
\int_{-\infty}^{+\infty}
\mathrm{d}t'\,
\left[
-\delta'\left(r-v(t-t')\right)
+
\frac{v(t-t')}{2}
\delta''\left(r-v(t-t')\right)
\right]
F_{ij}(t') .
\end{equation}
The first contribution is
\begin{equation}
\int_{-\infty}^{+\infty}
\mathrm{d}t'\,
\delta'\left[
r-v(t-t')
\right]
F_{ij}(t')
=
-\frac{1}{v^2}F_{ij}'(t_r)
=
-\frac{1}{2v^2}I_{ij}^{(3)}(t_r).
\end{equation}
Then,
\begin{equation}
-\int_{-\infty}^{+\infty}
\mathrm{d}t'\,
\delta'\left[
r-v(t-t')
\right]
F_{ij}(t')
=
\frac{1}{2v^2}I_{ij}^{(3)}(t_r).
\end{equation}
The second contribution is
\begin{equation}
\begin{aligned}
&\int_{-\infty}^{+\infty}
\mathrm{d}t'\,
\frac{v(t-t')}{2}
\delta''\left[
r-v(t-t')
\right]
F_{ij}(t') =
-\frac{1}{2v^2}I_{ij}^{(3)}(t_r)
+
\frac{r}{4v^3}I_{ij}^{(4)}(t_r).
\end{aligned}
\end{equation}
Therefore, the third--derivative terms cancel exactly so that we have:
\begin{equation}
\mathcal{C}_{ij}
=
\frac{r}{4v^3}I_{ij}^{(4)}(t_r).
\end{equation}
As a result, the complete helicity--resolved strain becomes
\begin{equation}
\label{strainnn}
h_{ij}^{(\lambda)}(t,{\bf r})
=
\frac{2G}{r}
\frac{I_{ij}^{(2)}(t_r)}{v^2}
+
\frac{2iG\lambda \mathring{k}^{(5)}_{(V)}}{v^3}
I_{ij}^{(4)}(t_r).
\end{equation}
Equivalently, this may be written as
\begin{equation}
\label{strainnnfactored}
h_{ij}^{(\lambda)}(t,{\bf r})
=
\frac{2G}{r}
\left[
\frac{1}{v^2}I_{ij}^{(2)}(t_r)
+
\frac{i\lambda \mathring{k}^{(5)}_{(V)} r}{v^3}
I_{ij}^{(4)}(t_r)
\right].
\end{equation}

Several points are worthy to be pointed out. First, the term controlled by $\mathring{k}^{(4)}_{(I)}$ does not generate higher time derivatives of the quadrupole, as shown recently in Ref. \cite{AraujoFilho:2026zyt}. However, it shifts the retarded time and rescales the waveform by $1/v^2$. Second, the CPT--odd coefficient $\mathring{k}^{(5)}_{(V)}$ produces a helicity--dependent correction proportional to $I_{ij}^{(4)}(t_r)$. The apparent $I_{ij}^{(3)}(t_r)$ contributions generated separately by the $\delta'$ and $\delta''$ terms cancel once the full Green function is convolved with the quadrupole source. Third, the factor $i\lambda$ is a consequence of working in the helicity basis. It signals birefringent propagation between the two circular polarizations. After recombining the helicity modes into the usual linear polarizations, this factor corresponds to a mixing between the $+$ and $\times$ sectors, not to an unphysical imaginary observable.

To make the last point explicit, we now rewrite the result in terms of the usual linear polarization basis. We use the circular polarization tensors
\begin{equation}
e_{ij}^{(\lambda)}
=
\frac{1}{\sqrt{2}}
\bigg(
e_{ij}^{+}
+
i\lambda e_{ij}^{\times}
\bigg) ,
\end{equation}
and the corresponding helicity components of the quadrupole,
\begin{equation}
I^{(\lambda)}(t_r)
=
\frac{1}{\sqrt{2}}
\bigg[
I_{+}(t_r)
-
i\lambda I_{\times}(t_r)
\bigg].
\end{equation}
Therefore, Eq.~\eqref{strainnnfactored} may be expressed in the compact
helicity form
\begin{equation}
h^{(\lambda)}(t,r)
=
\frac{2G}{r}
\left[
\frac{1}{v^2}
I^{(\lambda)(2)}(t_r)
+
\frac{i\lambda \mathring{k}^{(5)}_{(V)}r}{v^3}
I^{(\lambda)(4)}(t_r)
\right].
\end{equation}
The physical metric perturbation is obtained by summing the two helicity
contributions,
\begin{equation}
h_{ij}(t,r)
=
\sum_{\lambda=\pm 1}
h^{(\lambda)}(t,r)e_{ij}^{(\lambda)}
=
h_{+}(t,r)e_{ij}^{+}
+
h_{\times}(t,r)e_{ij}^{\times}.
\end{equation}
After carrying out this recombination, the observable linear polarizations
are found to be
\begin{equation}
h_{+}(t,r)
=
\frac{2G}{r}
\left[
\frac{1}{v^2}
I_{+}^{(2)}(t_r)
+
\frac{\mathring{k}^{(5)}_{(V)}r}{v^3}
I_{\times}^{(4)}(t_r)
\right],
\label{60}
\end{equation}
and
\begin{equation}
h_{\times}(t,r)
=
\frac{2G}{r}
\left[
\frac{1}{v^2}
I_{\times}^{(2)}(t_r)
-
\frac{\mathring{k}^{(5)}_{(V)}r}{v^3}
I_{+}^{(4)}(t_r)
\right].
\label{61}
\end{equation}
Equivalently, the full tensor waveform can be written as
\begin{equation}
\begin{aligned}
h_{ij}(t,r)
=
\frac{2G}{r}
\Bigg\{
&
\left[
\frac{1}{v^2}
I_{+}^{(2)}(t_r)
+
\frac{\mathring{k}^{(5)}_{(V)}r}{v^3}
I_{\times}^{(4)}(t_r)
\right]e_{ij}^{+} +
\left[
\frac{1}{v^2}
I_{\times}^{(2)}(t_r)
-
\frac{\mathring{k}^{(5)}_{(V)}r}{v^3}
I_{+}^{(4)}(t_r)
\right]e_{ij}^{\times}
\Bigg\},
\label{62}
\end{aligned}
\end{equation}
or, by substituting $r=v(t-t_r)$, we obtain
\begin{equation}
\begin{aligned}
h_{ij}(t,r)
=
\frac{2G}{v(t-t_r)}
\bigg\{
&
\left[
\frac{1}{v^2}
I_{+}^{(2)}(t_r)
+
\frac{\mathring{k}^{(5)}_{(V)}}{v^2}
(t-t_r)
I_{\times}^{(4)}(t_r)
\right]e_{ij}^{+}
\\
&
+
\left[
\frac{1}{v^2}
I_{\times}^{(2)}(t_r)
-
\frac{\mathring{k}^{(5)}_{(V)}}{v^2}
(t-t_r)
I_{+}^{(4)}(t_r)
\right]e_{ij}^{\times}
\bigg\}.
\end{aligned}
\end{equation}
These expressions are real. Therefore, the imaginary factor appearing in the helicity--resolved Green function does not make the strain complex. Its role is to rotate the two linear polarizations into each other: the $+$ mode receives a correction proportional to the fourth derivative of the $\times$ quadrupole component, while the $\times$ mode receives a correction with the opposite sign proportional to the fourth derivative of the $+$ component. In other words, the coefficient $\mathring{k}^{(5)}_{(V)}$ produces a helicity--dependent birefringent correction that becomes a polarization mixing effect in the real linear basis.

}

%%%%%%%%%%%%%%%%%%%%%%%%%%%%%%%%%%%%%%%%%%%%%%%%%%%%%%%%%%%%%%%%%%%%%%%%%%%%%%%%%%%%%%%%%%%%%%%%%%%%%%%%%%%%%%%%%%%%%%%%%%%%%%%%%%%%%%%%%%%%%%%%%%%%%%%%%%%%%%%%%%%%%%%%%%%%%%%%%%%%%%%%%%%%%%%%%%%%%%%%%%%%%%%%%%%%%%%%%%%%%%%%%%%%%%%%%%%%%%%%%%%%%%%%%%%%%%%%%%%%%%%%%%%%%%%%%%%%%%%%%%%%%%%%%%%%%%%%%%%%%%%%%%%%%%%%%%%%%%%%%%%%%%%%%%%%%%%%

\subsection{Application with a binary black hole system}

\begin{figure}
    \centering
    \includegraphics[scale=0.5]{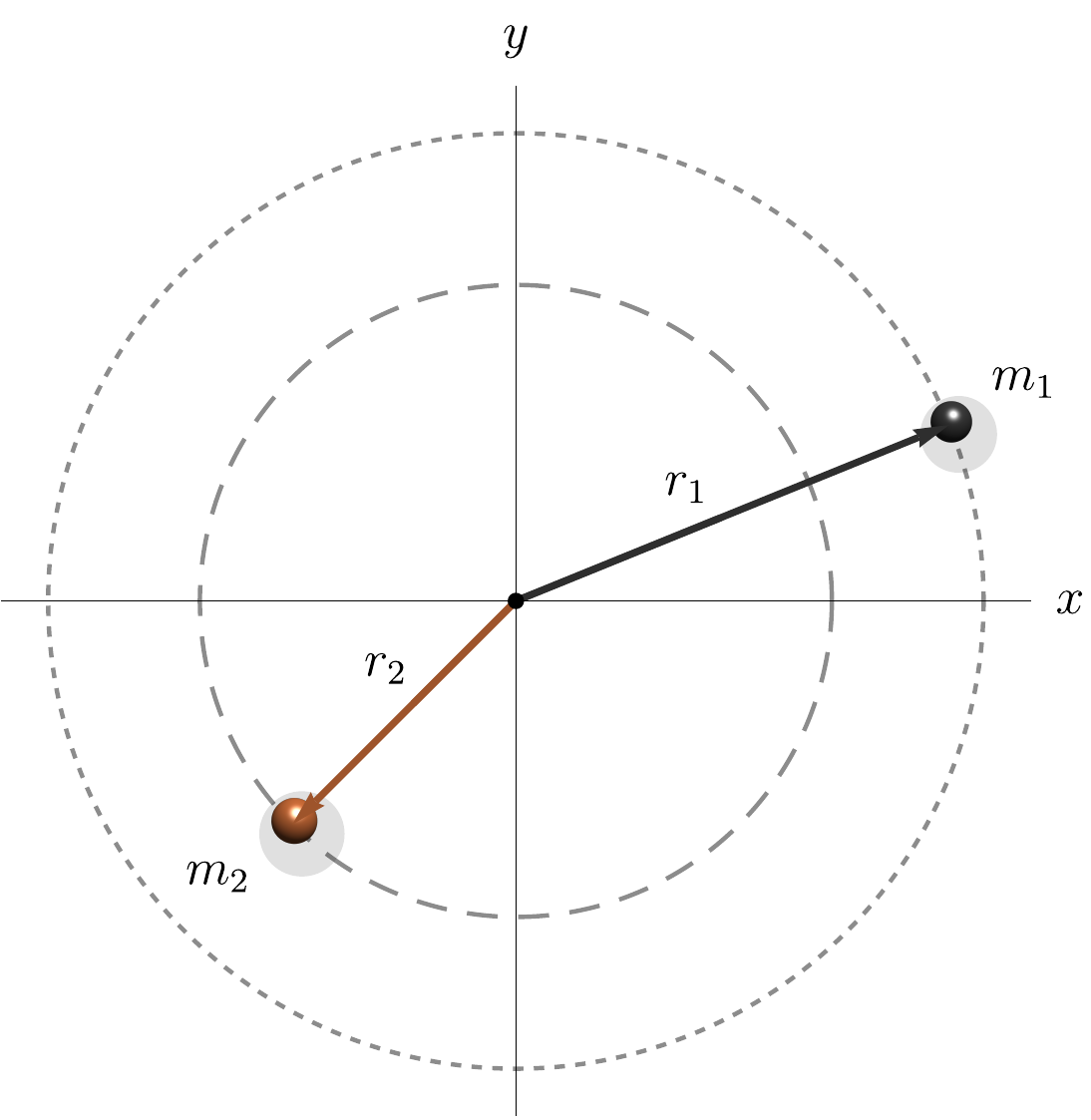}
    \caption{Illustration of a two-body black-hole system viewed in the barycentric frame. The compact constituents, labeled by masses $m_1$ and $m_2$, undergo orbital motion restricted to the $xy$ plane, with their trajectories specified by radii $r_1$ and $r_2$.}
    \label{binary}
\end{figure}

To illustrate these effects in a physical setting, we apply the formalism to a binary black hole system, schematically shown in Fig.~\ref{binary}. The configuration comprises two compact masses, $m_1$ and $m_2$, undergoing orbital motion restricted to the $xy$ plane. Adopting the center--of--mass frame, the motion of each body is characterized by its orbital radius, labeled $r_1$ and $r_2$, which specify their trajectories about the common barycenter.

The gravitational radiation is generated by a simplified source model in which the system is represented by two pointlike bodies undergoing planar orbital motion. The influence of this matter configuration on the spacetime geometry is captured through the associated stress--energy tensor, whose explicit form is given by
\begin{equation}
T_{00}
=
\delta(z)\Big[m_{1}\,\delta(x-x_{1})\delta(y-y_{1})
+
m_{2}\,\delta(x-x_{2})\delta(y-y_{2})\Big].
\end{equation}

The binary dynamics are modeled by assuming that each constituent executes uniform circular motion around the system’s barycenter. Within this approximation, the trajectories of the two bodies are specified by the relations
\begin{align}\label{eqmovbin}
x_{1}(t)=\frac{m_{2}l_{0}}{M}\cos(\omega t),
&\qquad
y_{1}(t)=\frac{m_{2}l_{0}}{M}\sin(\omega t), \nonumber\\
x_{2}(t)=-\frac{m_{1}l_{0}}{M}\cos(\omega t),
&\qquad
y_{2}(t)=-\frac{m_{1}l_{0}}{M}\sin(\omega t).
\end{align}
Notice that, in this notation, the total mass of the system is $M=m_1+m_2$, while the characteristic separation is given by $l_0=r_1+r_2$. The motion is assumed to be circular with angular velocity $\omega$, a choice that greatly simplifies the form of the mass quadrupole tensor. Under these assumptions, only three of its components remain nonvanishing,
\begin{align}\label{compI}
I_{xx}(t)
&=
\frac{\mu}{2}l_{0}^{2}\big[1+\cos(2\omega t)\big], \nonumber\\
I_{yy}(t)
&=
\frac{\mu}{2}l_{0}^{2}\big[1-\cos(2\omega t)\big], \nonumber\\
I_{xy}(t)
=
I_{yx}(t)
&=
\frac{\mu}{2}l_{0}^{2}\sin(2\omega t),
\end{align}
where we have introduced the quantity the reduced mass $\mu=m_1m_2/(m_1+m_2)$.

{Inserting the explicit expressions for the quadrupole moments given in Eq.~\eqref{compI}
into the real linear basis waveform of Eq.~\eqref{60}--\eqref{62}, after recombining the
helicity modes, one finds that only a subset of the metric perturbation components
survives in the radiation zone. The nonvanishing components are
\begin{align}
h_{xx}(t,r)
&=
-\frac{4G\,\mu\,l_{0}^{2}\,\omega^{2}}{r\,v^{2}}\,
\cos\!\big(2\omega\,t_{r}\big)
+
\frac{16G\,\mu\,l_{0}^{2}\,\omega^{4}}{v^{3}}\,
\mathring{k}^{(5)}_{(V)}
\sin\!\big(2\omega\,t_{r}\big),
\label{hxx}
\\
h_{yy}(t,r)
&=
+\frac{4G\,\mu\,l_{0}^{2}\,\omega^{2}}{r\,v^{2}}\,
\cos\!\big(2\omega\,t_{r}\big)
-
\frac{16G\,\mu\,l_{0}^{2}\,\omega^{4}}{v^{3}}\,
\mathring{k}^{(5)}_{(V)}
\sin\!\big(2\omega\,t_{r}\big),
\label{hyy}
\\
h_{xy}(t,r)
=
h_{yx}(t,r)
&=
-\frac{4G\,\mu\,l_{0}^{2}\,\omega^{2}}{r\,v^{2}}\,
\sin\!\big(2\omega\,t_{r}\big)
-
\frac{16G\,\mu\,l_{0}^{2}\,\omega^{4}}{v^{3}}\,
\mathring{k}^{(5)}_{(V)}
\cos\!\big(2\omega\,t_{r}\big).
\label{hxy}
\end{align}

}

These results make clear that the emitted signal retains the familiar tensorial structure associated with the quadrupole formula, even in the presence of Lorentz violation. The deviations instead arise through changes in how the waves propagate. One modification is controlled by the coefficient $\mathring{k}^{(4)}_{(I)}$, which alters the effective propagation speed $v$, thereby shifting the retarded time at which the waveform is evaluated, as shown in the previous augmentations\footnote{{As recently shown in the literature \cite{AraujoFilho:2026zyt}, the coefficient $\mathring{k}^{(4)}_{(I)}$ alone, we might say, does not generate a ``genuine'' correction to the gravitational wave amplitude. Its contribution appears only through an overall factor proportional to $1/v^{2}$, which can be absorbed into the gravitational coupling by means of the redefinition $G\to\widetilde{G}\equiv G/v^{2}$, as discussed in the present paper}; its effect is encoded through the retarded--time shift. }. The same coefficient also induces a uniform rescaling of the strain amplitude by a factor $1/v^{2}$, originating from the normalization used here. A second, qualitatively different effect is produced by the CPT--odd coefficient $\mathring{k}^{(5)}_{(V)}$. This term generates helicity--dependent corrections that enter through higher--order time derivatives of the mass quadrupole, reshaping the time dependence of the waveform while leaving the number of propagating degrees of freedom unchanged.

\begin{figure}
    \centering
    \includegraphics[scale=0.6]{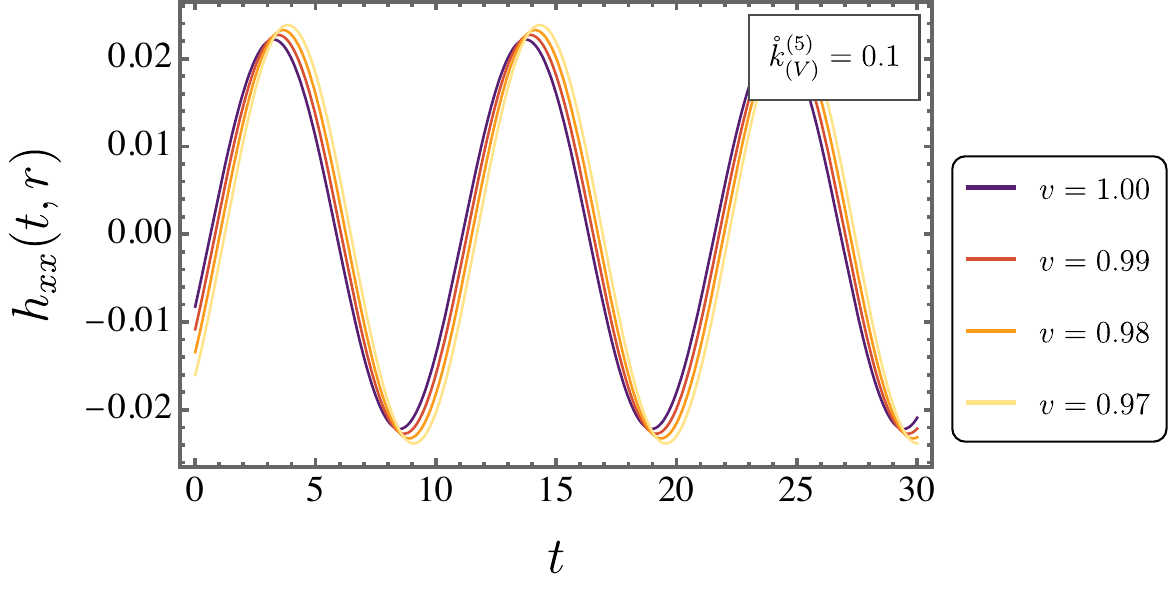}
    \caption{Waveform $h_{xx}(t,r)$ as a function of time $t$ for a representative configuration with $\omega=0.3$, $r=20$, $\mu=1$, and $l_0=1$. The signal incorporates the combined effects of the Lorentz--violating coefficients $\mathring{k}^{(4)}_{(I)}$ and $\mathring{k}^{(5)}_{(V)}$.}
    \label{hxx_waveform}
\end{figure}

{In Fig.~\ref{hxx_waveform}, we display the waveform $h^{+}_{xx}(t,r)$ as a function of the time $t$ for a representative configuration of the system, with $\omega=0.3$, $r=20$, $\mu=1$, and $l_0=1$. In this plot, the CPT--odd coefficient is fixed at $\mathring{k}^{(5)}_{(V)}=0.1$, while different values of the propagation parameter $v$ are considered. Since $v=1-\mathring{k}^{(4)}_{(I)}$, the parameter $v$ also encodes the contribution of the isotropic CPT--even coefficient $\mathring{k}^{(4)}_{(I)}$. In other words, we see that how the waveform are changed when the CPT--even and CPT--odd corrections are taken into account. The parameter $v$ affects the signal through the retarded time $t_r=t-r/v$, producing a shift in the oscillatory phase. It also appears in the amplitude prefactors, unless the corresponding overall contribution is absorbed into an effective gravitational coupling. In particular, since $t-t_r=r/v$, the first term of $h_{xx}(t,r)$ is effectively weighted by $1/(r v^2)$, whereas the term proportional to $\mathring{k}^{(5)}_{(V)}$ carries an additional factor $1/v^3$. The case $v=1$ corresponds to vanishing $\mathring{k}^{(4)}_{(I)}$, but it is still not the pure general--relativistic waveform because $\mathring{k}^{(5)}_{(V)}$ remains nonzero in all curves shown.}

\begin{figure}
    \centering
    \includegraphics[scale=0.6]{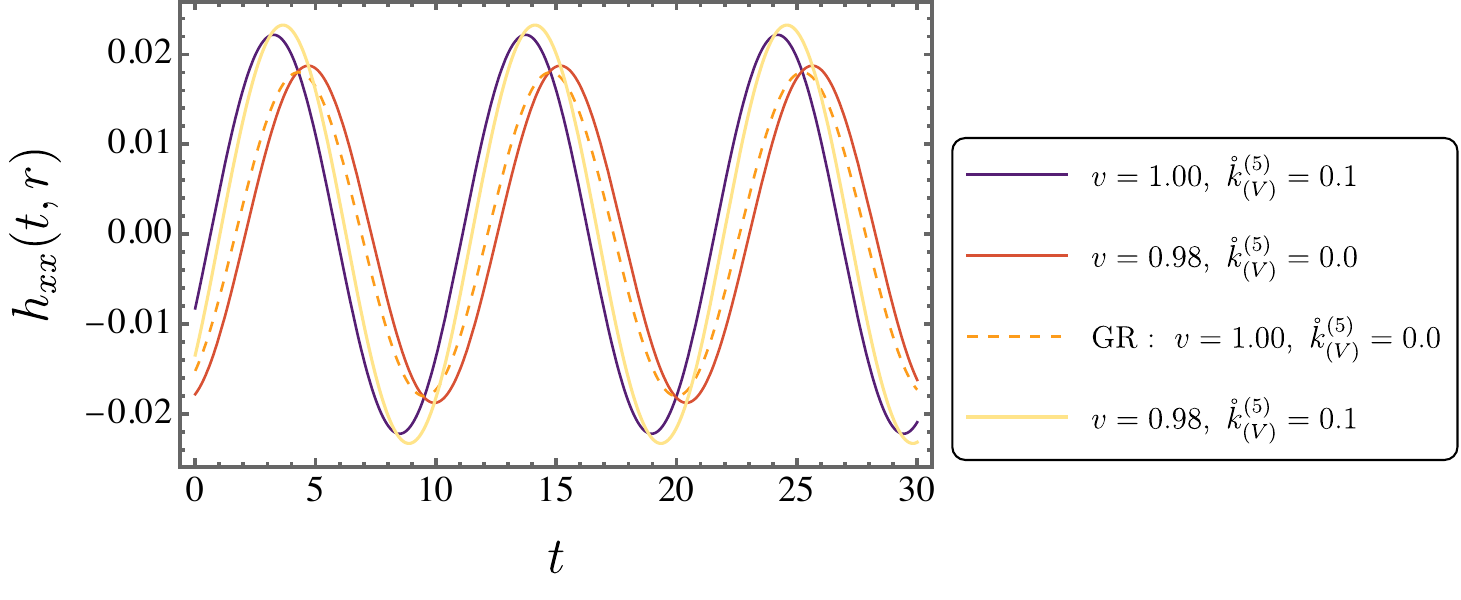}
    \caption{Impact of the Lorentz--violating operators $\mathring{k}^{(4)}_{(I)}$ and $\mathring{k}^{(5)}_{(V)}$ {on the waveform behavior in comparison with the general relativity case.}}
    \label{hxxfork}
\end{figure}

{In Fig.~\ref{hxxfork}, we compare the behavior of the waveform $h^{+}_{xx}(t,r)$ for four representative cases: the general--relativistic limit, the contribution of $\mathring{k}^{(4)}_{(I)}$ alone through $v\neq 1$, the contribution of $\mathring{k}^{(5)}_{(V)}$ alone, and the mixed case in which both effects are present. This figure provides a complementary comparison to Fig.~\ref{hxx_waveform}, since it separates the individual and combined roles of the two Lorentz--violating coefficients. The dashed curve corresponds to the general--relativistic reference, obtained for $v=1$ and $\mathring{k}^{(5)}_{(V)}=0$. When only $\mathring{k}^{(4)}_{(I)}$ is present, represented by $v=0.98$ and $\mathring{k}^{(5)}_{(V)}=0$, the waveform is shifted with respect to the GR curve due to the modified retarded time $t_r=t-r/v$. In the adopted parametrization, $v=1-\mathring{k}^{(4)}_{(I)}$, so the propagation parameter $v$ carries the contribution of the CPT--even coefficient $\mathring{k}^{(4)}_{(I)}$. In addition, if the corresponding overall factor is not absorbed into the effective coupling $\widetilde{G}$, the powers of $v$ in the prefactors also rescale the amplitude.  For $v=1$ and $\mathring{k}^{(5)}_{(V)}=0.1$, the deviation from the GR curve is caused solely by the CPT--odd correction, which introduces an additional oscillatory contribution proportional to $\sin(2\omega t_r)$. Since this term is in quadrature with the GR--like contribution proportional to $\cos(2\omega t_r)$, it modifies the relative phase and amplitude of the waveform. Finally, the mixed case, with $v=0.98$ and $\mathring{k}^{(5)}_{(V)}=0.1$, combines the phase shift generated by the modified retarded time with the extra CPT--odd oscillatory component.}

\begin{figure}
    \centering
    \includegraphics[scale=0.55]{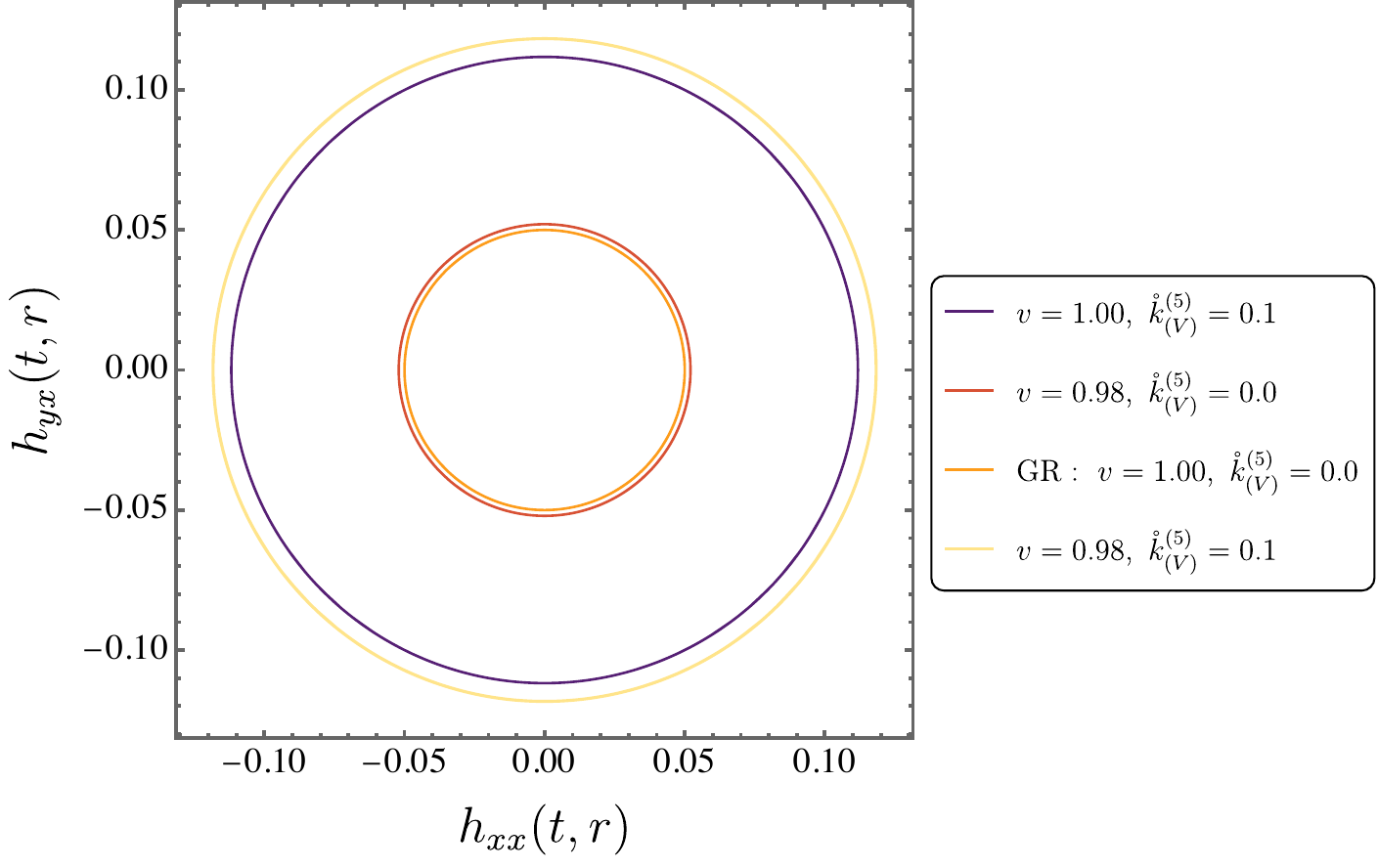}
    \caption{{Parametric behavior of the waveform in the $(h_{xx},h_{yx})$ plane for representative configurations. The GR case gives the reference circular trajectory. When only $\mathring{k}^{(4)}_{(I)}$ is present, the curve remains circular, with its radius modified by the explicit $1/v^2$ prefactor adopted in our normalization. When $\mathring{k}^{(5)}_{(V)}$ is switched on, the CPT--odd contribution enters in quadrature with the leading oscillatory terms, changing the radius of the closed trajectory.}}
    \label{paramtetricplotss}
\end{figure}

{In Fig.~\ref{paramtetricplotss}, we show the parametric behavior of the waveform in the
$(h_{xx},h_{yx})$ plane for four representative configurations. The GR reference,
$v=1.00$ and $\mathring{k}^{(5)}_{(V)}=0$, gives the standard circular trajectory. When
only $\mathring{k}^{(4)}_{(I)}$ is present, namely $v=0.98$ and
$\mathring{k}^{(5)}_{(V)}=0$, the curve remains circular, but its radius is modified by
the factor $1/v^{2}$ appearing in the leading quadrupole contribution. For
$v=1.00$ and $\mathring{k}^{(5)}_{(V)}=0.1$, the CPT--odd correction adds a component
in quadrature with the GR--like term, increasing the radius without changing the
closed form of the trajectory. In the mixed case, $v=0.98$ and
$\mathring{k}^{(5)}_{(V)}=0.1$, the two amplitude effects combine: the leading term is
weighted by $1/v^{2}$, while the CPT--odd contribution carries the factor $1/v^{3}$.}

\begin{figure}
    \centering
    \includegraphics[scale=0.65]{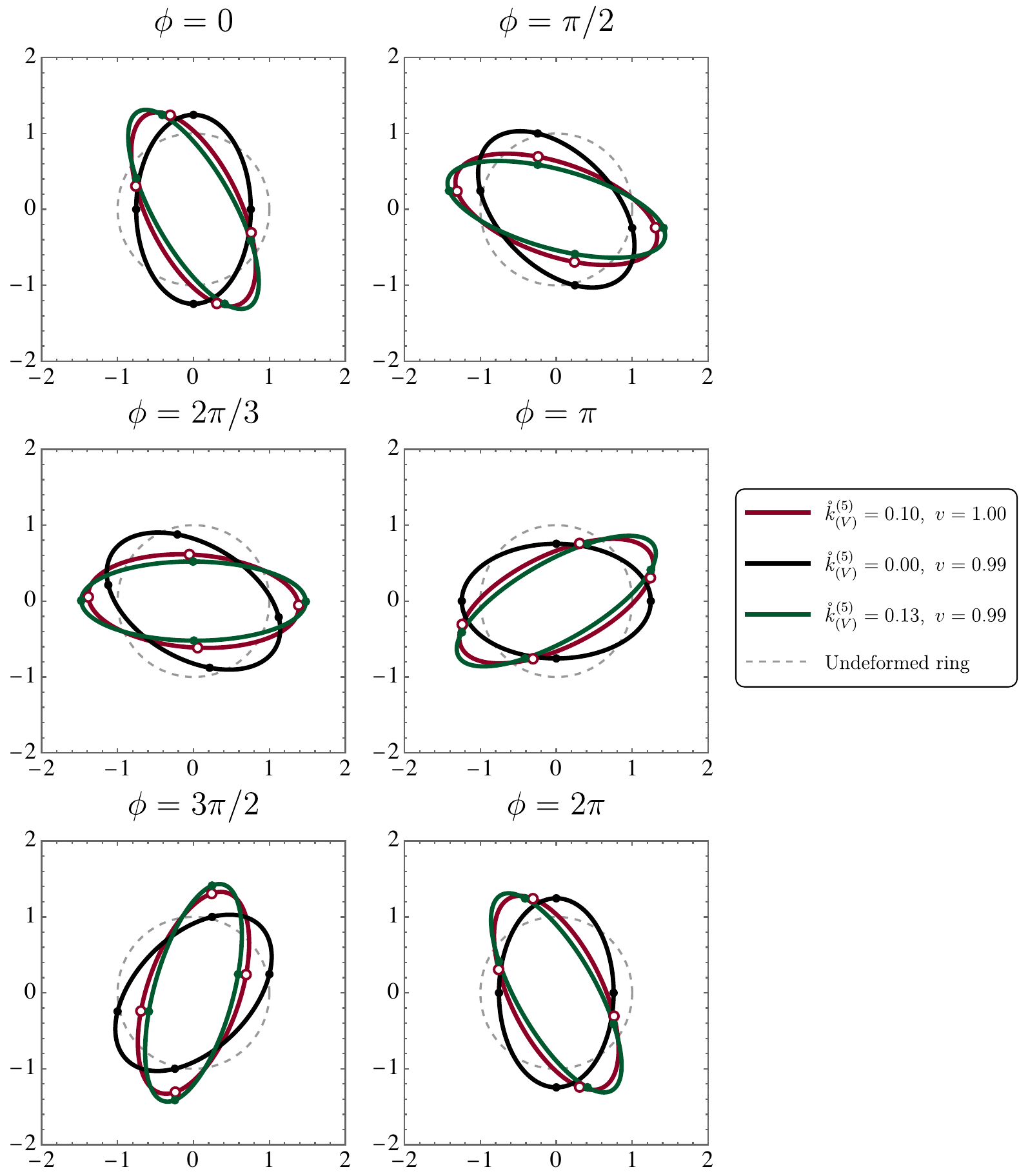}
    \caption{Transverse deformation of a ring of freely falling test particles induced by a gravitational wave.  Snapshots are shown at fixed phase $\phi=2\omega t_r$, with $t_r=t-r/v$.  The dashed gray circle is the undeformed ring.  The wine, black, and dark--green curves correspond, respectively, to $\mathring{k}^{(5)}_{(V)}=0.10$, $v=1.00$; $\mathring{k}^{(5)}_{(V)}=0$, $v=0.99$; and $\mathring{k}^{(5)}_{(V)}=0.13$, $v=0.99$.  The markers follow the same color coding and indicate selected particles along the principal axes.  We use $\mu=1$, $l_0=1$, $\omega=0.4$, and $r=20$.}
    \label{squeeze}
\end{figure}

Figure~\ref{squeeze} displays the transverse deformation of a ring of freely falling test particles produced by a gravitational wave propagating along the orthogonal direction. Each panel corresponds to a fixed value of the retarded phase $\phi = 2\omega t_r$, where $t_r = t - r/v$, and the sequence of snapshots spans one complete oscillation period.

{In all panels, the dashed gray circle represents the undeformed ring in the absence of the gravitational wave. The solid curves show the instantaneous displacement of the particles obtained from the linearized transverse--traceless mapping $x^i \rightarrow x^i + \tfrac{1}{2} h^{i}{}{j} x^j$, with nonvanishing strain components $h_{xx}$ and $h_{yx}$. The small markers indicate selected test particles initially placed along the principal axes of the ring, making the stretching and squeezing directions explicit at each phase. The wine curve, with open markers, corresponds to the case $\mathring{k}^{(5)}_{(V)} = 0.10$ and $v = 1.00$; the black curve, with black dots, corresponds to $\mathring{k}^{(5)}_{(V)} = 0$ and $v = 0.99$; and the dark--green curve, with filled markers, represents the combined case $\mathring{k}^{(5)}_{(V)} = 0.13$ and $v = 0.99$. The parameters used in the plots are $\mu = 1$, $l_0 = 1$, $\omega = 0.4$, and $r = 20$. As the phase evolves, the ring is deformed into ellipselike configurations whose orientation and eccentricity vary over the cycle, thereby illustrating how the $\mathring{k}^{(4)}_{(I)}$ and $\mathring{k}^{(5)}_{(V)}$ contributions modify the polarization pattern.}

%%%%%%%%%%%%%%%%%%%%%%%%%%%%%%%%%%%%%%%%%%%%%%%%%%%%%%%%%%%%%%%%%%%%%%%%%%%%%%%%%%%%%%%%%%%%%%%%%%%%%%%%%%%%%%%%%%%%%%%%%%%%%%%%%%%%%%%%%%%%%%%%%%%%%%%%%%%%%%%%%%%%%%%%%%%%%%%%%%%%%%%%%%%%%%%%%%%%%%%%%%%%%%%%%%%%%%%%%%%%%%%%%%%%%%%%%%%%%%%%%%%%%%%%%%%%%%%%%%%%%%%%%%%%%%%%%%%%%%%%%%%%%%%%%%%%%%%%%%%%%%%%%%%%%%%%%%%%%%%%%%%%%%%%%%%%%%%%%%%%%%%%%%%%%%%%%%%%%%%%%%%%%%%%%%%%%%%%%%%%%%%%%%%%

\section{Bounds from LVK (Virgo/KAGRA-including catalogs) data}
\label{sec:bounds_LVK}

This section updates the phenomenological bounds on the isotropic propagation coefficients by using \emph{published constraints inferred from LIGO--Virgo--KAGRA (LVK) analyses}. We emphasize that we do not reanalyze the strain data here; instead, we translate the LVK posterior constraints reported in the official catalogs and ``Tests of General Relativity'' analyses into bounds on $\mathring{k}^{(4)}_{(I)}$ and $\mathring{k}^{(5)}_{(V)}$ \cite{Abbott:2016blz,Abbott:2017oio,Abbott:2021gyp}. The numerical estimates presented below are based on the published GWTC--3 bounds on modified gravitational wave propagation.

%%%%%%%%%%%%%%%%%%%%%%%%%%%%%%%%%%%%%%%%%%%%%%%%%%%%%%%%%%%%%%%%%%%%%%%%%%%%%%%%%%%%%%%%%%%%%%%%%%%%%%%%%%%%%%%%%%%%%%%%%%%%%%%%%%%%%%%%%%%%%%%%
\subsection{Bound on the coefficient $\mathring{k}^{(4)}_{(I)}$}

In our isotropic parametrization, the coefficient is encoded through
\begin{equation}
v \equiv 1-\mathring{k}^{(4)}_{(I)},
\qquad\Rightarrow\qquad \frac{v_{\rm GW}}{c}
\equiv
1-\mathring{k}^{(4)}_{(I)},
\qquad
|\mathring{k}^{(4)}_{(I)}|
=
\left|1-\frac{v_{\rm GW}}{c}\right|.
\end{equation}

A robust LVK bound comes from the multimessenger association GW170817/GRB~170817A, observed by the Advanced LIGO and Advanced Virgo detectors, which constrains the fractional difference between the GW speed and the speed of light to be \cite{Abbott:2017oio}
\begin{equation}
-3\times 10^{-15}
\;\lesssim\;
\frac{v-c}{c}
\;\lesssim\;
7\times 10^{-16}.
\end{equation}
Therefore, within the present isotropic model,
\begin{equation}
\left|\mathring{k}^{(4)}_{(I)}\right|
\;\lesssim\;
3\times 10^{-15}.
\label{eq:k4_bound_GW170817}
\end{equation}
For comparison, independent bounds obtained from detector-network timing using GW observations alone (without electromagnetic counterparts) yield significantly weaker constraints, typically $v\simeq(0.97\text{--}1.01)c$ when combining O1--O2 events \cite{Abbott:2016blz}, and therefore do not compete with the multimessenger limit.

%%%%%%%%%%%%%%%%%%%%%%%%%%%%%%%%%%%%%%%%%%%%%%%%%%%%%%%%%%%%%%%%%%%%%%%%%%%%%%%%%%%%%%%%%%%%%%%%%%%%%%%%%%%%%%%%%%%%%%%%%%%%%%%%%%%%%%%%%%%%%%%%
\subsection{Bounds on the CPT--odd $d=5$ coefficient $\mathring{k}^{(5)}_{(V)}$}

{
\subsubsection{Translation from LVK propagation tests}
}

To connect our $d=5$ CPT--odd term with LVK propagation tests, we employ the standard parameterized modified dispersion relation adopted in the LVK ``Tests of GR'' framework \cite{Abbott:2021gyp},
\begin{equation}
E^{2} = p^{2}c^{2} + A_{\alpha}\,p^{\alpha}c^{\alpha},
\label{eq:LVK_dispersion}
\end{equation}
where constraints on $A_{\alpha}$ are obtained by combining multiple events across the LVK catalogs.

For $\alpha=3$, Eq.~\eqref{eq:LVK_dispersion} yields (in units with $c=\hbar=1$)
\begin{equation}
\omega^{2} = p^{2} + A_{3}p^{3}
\qquad\Rightarrow\qquad
\omega \simeq p + \frac{1}{2}A_{3}p^{2},
\end{equation}
to leading order in $A_{3}$.

On the other hand, the helicity-resolved dispersion relation derived in this work takes the form
\begin{equation}
\omega_{\lambda}
\simeq
(1-\mathring{k}^{(4)}_{(I)})\,p
+
\lambda\,\mathring{k}^{(5)}_{(V)}\,p^{2},
\qquad \lambda=\pm1.
\label{eq:our_dispersion_linear}
\end{equation}
Neglecting the tiny overall rescaling induced by $\mathring{k}^{(4)}_{(I)}$ in this matching step, the LVK \emph{nonbirefringent} coefficient $A_{3}$ constrains the magnitude of any $p^{2}$ correction to the phase velocity. A conservative translation is therefore obtained by identifying the size of the $p^{2}$ term,
\begin{equation}
\left|\mathring{k}^{(5)}_{(V)}\right|
\;\lesssim\;
\frac{|A_{3}|}{2}.
\label{eq:k5_from_A3}
\end{equation}
This estimate is conservative, since the LVK $A_{3}$ analysis is not optimized for helicity--dependent birefringence; a dedicated birefringent waveform template would be expected to constrain $\mathring{k}^{(5)}_{(V)}$ more directly \cite{Kostelecky:2016kfm,Mewes:2019dhj}.

\paragraph{Numerical value from GWTC-3 propagation bounds.}

The LVK propagation tests reported in GWTC--3 constrain the dimensionless quantity $\bar{A}_{\alpha}\equiv A_{\alpha}/\mathrm{eV}^{\,2-\alpha}$ \cite{Abbott:2021gyp}. For $\alpha=3$, the combined 90\% credible bound reads
\begin{equation}
|\bar{A}_{3}|
\;\lesssim\;
\left(0.45\text{--}1.16\right)\times 10^{-2}.
\end{equation}
{Equivalently, since $A_3$ has dimensions of $\mathrm{eV}^{-1}$, we have}
\begin{equation}
|A_{3}|
\;\lesssim\;
\left(0.45\text{--}1.16\right)\times 10^{-2}\ \mathrm{eV}^{-1}.
\end{equation}

Using $1~\mathrm{eV}^{-1}\simeq 6.582\times 10^{-16}\ \mathrm{s}$, Eq.~\eqref{eq:k5_from_A3} yields
\begin{equation}
\left|\mathring{k}^{(5)}_{(V)}\right|
\;\lesssim\;
\left(1.5\text{--}3.8\right)\times 10^{-18}\ \mathrm{s},
\qquad (90\%~\mathrm{C.I.},\ \text{LVK GWTC-3 translation}).
\label{eq:k5_numeric_LVK}
\end{equation}

%%%%%%%%%%%%%%%%%%%%%%%%%%%%%%%%%%%%%%%%%%%%%%%%%%%%%%%%%%%%%%%%%%%%%%%%%%%%%%%%%%%%%%%%%%%%%%%%%%%%%%%%%%%%%%%%%%%%%%%%%%%%%%%%%%%%%%%%%%%%%%%%
\subsubsection{Birefringent phase accumulation}

An independent and physically transparent constraint on $\mathring{k}^{(5)}_{(V)}$ follows from the absence of large polarization rotation in observed GW signals. 
{Since the strain components in
Eqs.~(\ref{hxx})--(\ref{hxy}) depend on the phase $2\omega t_r$, the angular frequency of
the emitted gravitational wave is
$\Omega \equiv 2\omega = 2\pi f$.} The helicity--dependent {propagation law implies an accumulated relative phase}
\begin{equation}
\Delta\phi
=
{\big[p_-(\Omega)-p_+(\Omega)\big]L
\simeq
\frac{2\mathring{k}^{(5)}_{(V)}}{v}\frac{\Omega^2 L}{c}
=
\frac{8\mathring{k}^{(5)}_{(V)}}{v}\frac{\omega^2 L}{c}}.
\end{equation}
{This phase} induces mixing between the $(+,\times)$ polarization states during propagation \cite{Kostelecky:2016kfm}.

Requiring that the accumulated birefringent phase remains smaller than an ${\cal O}(1)$ tolerance,
\begin{equation}
|\Delta\phi|\lesssim 1,
\end{equation}
yields the conservative bound
\begin{equation}
\left|\mathring{k}^{(5)}_{(V)}\right|
\lesssim {
\frac{vc}{2\Omega^2L}
=
\frac{vc}{8\omega^2L}}.
\label{eq:k5_Deltaphi_bound}
\end{equation}

For representative LVK sources with $f\sim 100~\mathrm{Hz}$ and distances $L\sim 40$--$500~\mathrm{Mpc}$, this estimate {gives}
\begin{equation}
{
\left|\mathring{k}^{(5)}_{(V)}\right|
\lesssim
3.1\times 10^{-22}
\left(\frac{100~\mathrm{Hz}}{f}\right)^2
\left(\frac{40~\mathrm{Mpc}}{L}\right)
\,\mathrm{s}.
}
\end{equation}
{Across the interval $L\sim 40$--$500~\mathrm{Mpc}$, this corresponds to}
\begin{equation}
{
\left|\mathring{k}^{(5)}_{(V)}\right|
\lesssim
(2.5\times 10^{-23}\text{--}3.1\times 10^{-22})~\mathrm{s},
\qquad
f\sim 100~\mathrm{Hz}.
}
\end{equation}
{This phase-consistency estimate is numerically stronger than the catalog-level translation in Eq.~\eqref{eq:k5_numeric_LVK}; however, it should not be interpreted as a direct LVK catalog posterior.} Importantly, Eq.~\eqref{eq:k5_Deltaphi_bound} directly probes the distinctive birefringent signature of the CPT--odd $d=5$ operator and is closely tied to the regime of validity of the perturbative waveform expansion derived in this work.

%%%%%%%%%%%%%%%%%%%%%%%%%%%%%%%%%%%%%%%%%%%%%%%%%%%%%%%%%%%%%%%%%%%%%%%%%%%%%%%%%%%%%%%%%%%%%%%%%%%%%%%%%%%%%%%%%%%%%%%%%%%%%%%%%%%%%%%%%%%%%%%%

Collecting the results, a consistent set of bounds for the isotropic coefficients is
\begin{align}
\left|\mathring{k}^{(4)}_{(I)}\right|
\;\lesssim\;
3\times 10^{-15}
&\qquad \text{(GW170817/GRB~170817A; LV)},\\[4pt]
\left|\mathring{k}^{(5)}_{(V)}\right|
\;\lesssim\;
\left(1.5\text{--}3.8\right)\times 10^{-18}\ \mathrm{s}
&\qquad \text{(LVK GWTC--3 propagation tests)},\\[4pt]
\left|\mathring{k}^{(5)}_{(V)}\right|
\;\lesssim\;
{
\frac{v c}{2\Omega^{2}L}
=
\frac{v c}{8\omega^{2}L}
\simeq
\frac{c}{2\Omega^{2}L}
=
\frac{c}{8\omega^{2}L}}
&\qquad \text{(birefringent phase consistency bound)}.
\end{align}
{Using $r=L/c$ and $\Omega=2\omega$, the strain--control parameter associated with Eqs.~(67)--(69) can be written as}
\begin{equation}
\eta_5 {
=
4\left|\mathring{k}^{(5)}_{(V)}\right|
\frac{L}{c}\frac{\omega^2}{v}
=
\frac{|\Delta\phi|}{2}}.
\end{equation}

{These} bounds can be inserted directly into {the} waveform expressions and into the polarization--mixing phase $\Delta\phi$ to verify that the perturbative regime {remains under control}. 

%%%%%%%%%%%%%%%%%%%%%%%%%%%%%%%%%%%%%%%%%%%%%%%%%%%%%%%%%%%%%%%%%%%%%%%%%%%%%%%%%%%%%%%%%%%%%%%%%%%%%%%%%%%%%%%%%%%%%%%%%%%%%%%%%%%%%%%%%%%%%%%%
\subsubsection{Strain--based effects}
\label{subsubsec:strain_bounds}

The bounds above rely on catalog-level propagation analyses and on phase-accumulation arguments. One can also obtain complementary \emph{strain-based} consistency bounds by using the fact that LVK observations show that the measured strain is well described by GR templates, with no statistically significant evidence for {non--GR polarization anomalies}. In what follows we do not reanalyze detector data; instead, we impose conservative requirements directly on the analytic strain components \eqref{hxx}--\eqref{hxy}.

\paragraph{{Term control in the linear-basis strain.}}
{
From Eqs.~\eqref{hxx}--\eqref{hxy}, the corrected $d=5$ contribution no longer contains a secular term proportional to $t_r\omega^{4}$. Instead, the CPT--odd correction appears as a quadrature contribution to the leading GR-like oscillatory structure. Defining
}
\begin{equation}
{
A=\frac{4G\mu l_0^2\omega^2}{r v^2},
\qquad
B=\frac{16G\mu l_0^2\omega^4}{v^3}\mathring{k}^{(5)}_{(V)},
\qquad
\phi=2\omega t_r ,
}
\end{equation}
{
the relevant components can be written as
}
\begin{equation}
{
h_{xx}=-A\cos\phi+B\sin\phi,
\qquad
h_{yx}=-A\sin\phi-B\cos\phi .
}
\end{equation}
{
Thus, instead of using a component-wise ratio, which becomes ill-defined near the zeros of the leading oscillatory term, we impose the perturbative condition on the polarization-plane norm,
}
\begin{equation}
{
\eta_5
\equiv
\frac{\sqrt{(h^{(5)}_{xx})^2+(h^{(5)}_{yx})^2}}
{\sqrt{(h^{(0)}_{xx})^2+(h^{(0)}_{yx})^2}}
=
\frac{|B|}{A}
=
4\,\left|\mathring{k}^{(5)}_{(V)}\right|
\frac{r\omega^2}{v}.
}
\label{eq:strain_ratio_criterion}
\end{equation}
{
Requiring $\eta_5\lesssim\epsilon$, with $\epsilon\ll1$, gives
}
\begin{equation}
{
\left|\mathring{k}^{(5)}_{(V)}\right|
\;\lesssim\;
\frac{\epsilon v}{4r\omega^{2}}
}
\qquad
{
\text{(strain consistency)}.
}
\label{eq:k5_bound__strain}
\end{equation}
{
Restoring units through $r=L/c$, where $L$ is the propagation distance, this estimate becomes
}
\begin{equation}
{
\left|\mathring{k}^{(5)}_{(V)}\right|
\;\lesssim\;
\frac{\epsilon v c}{4L\omega^{2}}
}
\qquad
{
\text{(strain consistency, SI units)}.
}
\end{equation}
{
The corrected strain-based estimate is therefore controlled by the propagation distance, not by the in-band duration $T$. This change follows directly from the absence of the previous secular contribution proportional to $t_r\omega^4$.
}

\paragraph{{SNR-based perturbative consistency estimate.}}
{
A second way to express the same perturbative requirement is to demand that the CPT--odd correction does not induce a fractional deformation larger than the statistical scale of the observed signal. Let $\rho$ denote the network signal-to-noise ratio. Taking $\epsilon\sim1/\rho$, Eq.~\eqref{eq:k5_bound__strain} yields
}
\begin{equation}
{
\left|\mathring{k}^{(5)}_{(V)}\right|
\;\lesssim\;
\frac{v}{4\rho\,r\,\omega^{2}}
}
\qquad
{
\text{(SNR-based strain consistency)}.
}
\label{eq:k5_bound_SNR}
\end{equation}
{
or, restoring $r=L/c$,
}
\begin{equation}
{
\left|\mathring{k}^{(5)}_{(V)}\right|
\;\lesssim\;
\frac{v c}{4\rho\,L\,\omega^{2}}
}
\qquad
{
\text{(SNR-based strain consistency, SI units)}.
}
\end{equation}
{
This expression replaces the former $T$-dependent estimate. In the corrected waveform, increasing the observation time alone does not generate an accumulated strain growth; the relevant accumulation is instead associated with propagation over the distance $L$.
}

%%%%%%%%%%%%%%%%%%%%%%%%%%%%%%%%%%%%%%%%%%%%%%%%%%%%%%%%%%%%%%%%%%%%%%%%%%%%%%%%%%%%%%%%%%%%%%%%%%%%%%%%%%%%%%%%%%%%%%%%%%%%%%%%%%%%%%%%%%%%%%%%

\paragraph{Numerical illustration using benchmark LVK events.}
{
Although we do not reanalyze the detector strain, it is useful to attach indicative numbers to the corrected strain-based estimates by adopting representative event characteristics reported by the LVK Collaboration. For GW150914, we use $\rho\simeq24$ and a representative luminosity distance $L\simeq410~{\rm Mpc}$ \cite{Abbott:2016blz}. For GW170817, we use $\rho\simeq32.4$ and $L\simeq40~{\rm Mpc}$ \cite{Abbott:2017oio}.
}

{
We adopt a representative gravitational-wave frequency $f\sim100~{\rm Hz}$, so that $\omega=\pi f$ in our notation, since the oscillatory terms in \eqref{hxx}--\eqref{hxy} involve $\cos(2\omega t_r)$ and $\sin(2\omega t_r)$. Setting $v\simeq1$ and using $1~{\rm Mpc}/c\simeq1.03\times10^{14}~{\rm s}$, Eq.~\eqref{eq:k5_bound_SNR} gives
}
\begin{align}
{
\left|\mathring{k}^{(5)}_{(V)}\right|
}
&{
\lesssim
\frac{c}{4\rho\,L\,\omega^{2}}
\;\sim\;
2.5\times10^{-24}\ {\rm s}
}
\nonumber\\
&{
\hspace{3.0cm}
\text{(GW150914-like: $\rho\simeq24$, $L\simeq410~{\rm Mpc}$, $f\simeq100~{\rm Hz}$)},
}
\\[4pt]
{
\left|\mathring{k}^{(5)}_{(V)}\right|
}
&{
\lesssim
\frac{c}{4\rho\,L\,\omega^{2}}
\;\sim\;
1.9\times10^{-23}\ {\rm s}
}
\nonumber\\
&{
\hspace{3.0cm}
\text{(GW170817-like: $\rho\simeq32.4$, $L\simeq40~{\rm Mpc}$, $f\simeq100~{\rm Hz}$)}.
}
\end{align}

%%%%%%%%%%%%%%%%%%%%%%%%%%%%%%%%%%%%%%%%%%%%%%%%%%%%%%%%%%%%%%%%%%%%%%%%%%%%%%%%%%%%%%%%%%%%%%%%%%%%%%%%%%%%%%%%%%%%%%%%%%%%%%%%%%%%%%%%%%%%%%%%%%%%%%%%%%%%%%%%%%%%%%%%%%%%%%%%%%%%%%%%%%%%%%%%%%%%%%%%%%%%%%%%%%%%%%%%%%%%%%%%%%%%%%%%%%%%%%%%%%%%%%%%%%%%%%%%%%%%%%%%%%%%%%%%%%%%%%%%%%%%%%%%%%%%%%%%%%%%%%%%%%%%%%%%%%%%%%%%%%%%%%%%%%%%%%%%%%%%%%%%%%%%%%%%%%%%%%%%%%%%%%%%%%%%%%%%%%%%%%%%%%%%

\section{Conclusion}

In this work, we analyzed how Lorentz-- and diffeomorphism--violating operators in the linearized gravitational sector of the Standard Model Extension modified the propagation of gravitational waves. Restricting attention to the transverse--traceless tensor sector and to isotropic coefficients, we focused on the combined effects of the CPT--even dimension--four coefficient $\mathring{k}^{(4)}_{(I)}$ and the CPT--odd dimension--five coefficient $\mathring{k}^{(5)}_{(V)}$.

Starting from the quadratic action for metric perturbations, we derived the modified dispersion relation and showed that Lorentz violation altered gravitational wave propagation through two distinct mechanisms. The coefficient $\mathring{k}^{(4)}_{(I)}$ rescaled the propagation speed, leading to a shift in the retarded time and an overall amplitude rescaling of the waveform. In contrast, the CPT--odd coefficient $\mathring{k}^{(5)}_{(V)}$ introduced helicity--dependent effects, giving rise to birefringence and polarization mixing without activating additional propagating degrees of freedom.

We constructed the retarded Green function associated with the modified wave operator and obtained explicit expressions for the gravitational waveform generated by matter sources. When applied to a binary black hole system, the resulting radiation preserved the standard quadrupole tensorial structure but acquired characteristic propagation--induced distortions. These included higher time--derivative corrections to the quadrupole formula, helicity--dependent phase shifted, {and there appeared a modification in} the strain amplitude.

{Using GW170817/GRB 170817A, published GWTC--3 propagation tests, polarization consistency arguments, and strain--based estimates, we translated existing observational results into bounds on the isotropic coefficients.} {The multimessenger event GW170817 constrained the isotropic CPT--even coefficient to $|\mathring{k}^{(4)}_{(I)}|\lesssim 3\times10^{-15}$. For the CPT--odd coefficient, the conservative translation of GWTC--3 modified--dispersion bounds yielded $|\mathring{k}^{(5)}_{(V)}|\lesssim (1.5-3.8)\times10^{-18}\,\mathrm{s}$, while birefringent phase--consistency and strain-control estimates reached the $10^{-23}$--$10^{-22}\,\mathrm{s}$ range for representative LVK frequencies and distances.}

As a further perspective, it would be worthwhile to examine the role of the remaining Lorentz--violating coefficients, for instance
$\mathring{k}^{(6)}_{(I)}$, $\mathring{k}^{(7)}_{(V)}$, and $\mathring{k}^{(8)}_{(I)}$, either separately or in suitable combinations. Such an analysis may uncover new or additional effects associated with birefringence and with the propagation properties of gravitational waves. Since these higher--dimension coefficients enter at progressively higher powers of the frequency $\omega$, identifying which ones are responsible for the expected ``unusual'' behavior is particularly relevant. In addition, exploring the modified dispersion relations derived here within the framework of ensemble theory appears to be a natural direction for future work, as discussed in Refs.~\cite{araujo2022does,araujo2022thermal,furtado2023thermal,oliveira2020thermodynamic,araujo2023thermodynamics}.

%%%%%%%%%%%%%%%%%%%%%%%%%%%%%%%%%%%%%%%%%%%%%%%%%%%%%%%%%%%%%%%%%%%%%%%%%%%%%%%%%%%%%%%%%%%%%%%%%%%%%%%%%%%%%%%%%%%%%%%%%%%%%%%%%%%%%%%%%%%%%%%%%%%%%%%%%%%%%%%%%%%%%%%%%%%%%%%%%%%%%%%%%%%%%%%%%%%%%%%%%%%%%%%%%%%%%%%%%%%%%%%%%%%%%%%%%%%%%%%%%%%%%%%%%%%%%%%%%%%%%%%%%%%%%%%%%%%%%%%%%%%%%%%%%%%%%%%%%%%%%%%%%%%%%%%%%%%%%%%%%%%%%%%%%%%%%%%%%%%%%%%%%%%%%%%%%%%%%%%%%%%%%%%%%%%%%%%%%%%%%%%%%%%%

\section{Acknowledgments}

\hspace{0.5cm}
A. A. Araújo Filho is supported by Conselho Nacional de Desenvolvimento Cient\'{\i}fico e Tecnol\'{o}gico (CNPq) and Fundação de Apoio à Pesquisa do Estado da Paraíba (FAPESQ), project numbers 150223/2025-0 and 1951/2025. N. H is supported by the Conselho Nacional de Desenvolvimento Científico e Tecnológico (CNPq), grant No. 152891/2025-0. N. H. also acknowledges the networking support provided by COST Action CA22113 -- Fundamental challenges in theoretical physics (Theory and Challenges), CA21106 -- COSMIC WISPers in the Dark Universe: Theory, astrophysics and experiments (CosmicWISPers), CA21136 -- Addressing observational tensions in cosmology with systematics and fundamental physics (CosmoVerse), and CA23130 -- Bridging high and low energies in search of quantum gravity (BridgeQG). I. P. L. was partially supported by the National Council for Scientific and Technological Development - CNPq, grant 312547/2023-4 and to acknowledge networking support by the COST Action BridgeQG (CA23130), the COST Action RQI (CA23115) and the COST Action FuSe (CA24101) supported by COST (European Cooperation in Science and Technology).  I. P. L. also acknowledges the discussions carried out in the VirgoBR group.

%%%%%%%%%%%%%%%%%%%%%%%%%%%%%%%%%%%%%%%%%%%%%%%%%%%%%%%%%%%%%%%%%%%%%%%%%%%%%%%%%%%%%%%%%%
\section{Data Availability Statement}

Data Availability Statement: No Data associated in the manuscript

%%%%%%%%%%%%%%%%%%%%%%%%%%%%%%%%%%%%%%%%%%%%%%%%%%%%%%%%%%%%%%%%%%%%%%%%%%%%%%%%%%%%%%%%%%

\bibliographystyle{ieeetr}
\bibliography{main}

\end{document}